\documentclass[11pt,a4paper]{article}
\usepackage{a4,hhline,array}
\usepackage[latin1]{inputenc}
\usepackage{epsfig,float}
\usepackage[page]{appendix}
\usepackage{graphicx, amsmath, amssymb}
\usepackage{hyperref}
\usepackage{url}
\usepackage[numbers]{natbib}
\usepackage{multibib,subfigure}
\usepackage{color,soul}
\usepackage{float}

\makeatletter
	\renewcommand{\subsubsection}{\@startsection{subsubsection}%
	{3}%
	{\z@}%
	{0.0ex plus 0.0ex minus 0.0ex}%
	{-2.0ex plus 0.0ex minus 0.0ex}%
	{\normalfont\normalsize\bfseries}}%
\makeatother
	\let\origitemize\itemize
	\def\itemize{\origitemize\itemsep-5pt}
	
\begin{document}

\pagestyle{empty}

\title{Technical Report: Artificial DNA - a Concept for Self-Building Embedded Systems}
\author{Uwe Brinkschulte\\
Institut für Informatik\\
Johann Wolfgang Goethe Universität Frankfurt, Germany\\
Email: brinks@es.cs.uni-frankfurt.de}

\maketitle
\thispagestyle{empty}

\section{Introduction}
\label{sec:introduction}
Embedded systems are growing more and more complex because of the increasing chip integration density, larger number of chips in distributed applications and demanding application fields (e.g. in cars and in households). In the near future it will become reality to have thousands of computing nodes within an embedded system. Bio-inspired techniques like self-organization are a key feature to handle this complexity. We have developed the Artificial Hormone System (AHS) as a decentralized, self-organizing, self-healing and self-optimizing mechanism to assign tasks to computing nodes of an embedded real-time system. The AHS is able to handle task assignment in complex embedded systems with a large number of processor cores.

However, to do so the AHS needs a blueprint of the structure and organization of the embedded application. This covers the segmentation of the application into tasks, the cooperation and communication between these tasks, the suitability of the processor cores for each of these tasks, etc. Currently, these assignments are done manually by the system developer, but in the future this is no longer feasable for large embedded systems having a large number of cores and tasks.

The idea is to follow again a bio-inspired principle. In biology the structure and organization of a system is coded in its DNA. This can be adopted to embedded systems. The blueprint of the structure and organization of the embedded system will be represented by an artificial DNA. The artificial DNA has to be be held compact and stored in every processor core of the system (like the biological DNA is stored in every cell of an organism). This makes the system \textit{self-describing}. Now, a mechanism like the AHS can transcribe the artificial DNA to setup and operate the embedded system accordingly at run-time. All the needed information for such a process like task structure, cooperation, communication and core suitability can be derived from the artificial DNA. Therefore, the system becomes \textit{self-building} at run-time based on its DNA. This enables a maximum amount of robustness, adaptivity and flexibility.

This technical report describes in detail the basic principles of the artificial DNA (Section \ref{sec:conception}) and its relationship to standard design methods for embedded systems (Section \ref{sec:relwork}). A prototypic implementation is presented (Section \ref{sec:implementation}) and evaluated (Section \ref{sec:evaluation}). Additionally, future work is detailly described (Section \ref{sec:futurework}) and a conclusion is given (Section \ref{sec:conclusions}). The technical report extends and complements the publications \cite{brinksCandC} and \cite{BrinksJournal2017}.

\section{Conception}
\label{sec:conception}
In the following, the basic conception of the proposed approach is explained in detail. It consists of the system composition model, the basic idea, the structure of the artificial DNA and how a system is built from its artificial DNA.

\subsection{System Composition Model}
\label{sec:compmod}
\begin{figure}[ht]
     \centering
	   \includegraphics[width=0.6\linewidth]{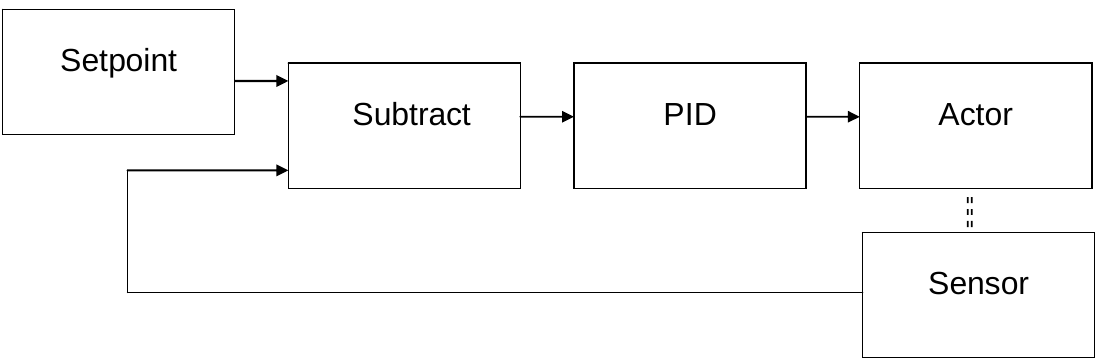}
	   \caption{Sample model of a simple embedded system}
	   \label{fig:controller}
\end{figure}

\begin{figure}[ht]
  \begin{minipage}[t]{0.45\linewidth}
	   \includegraphics[width=0.7\linewidth]{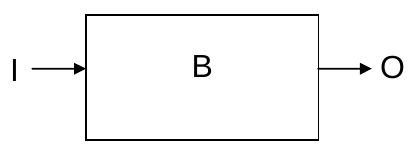}
	   \caption{General representation of a basic functional element}
	   \label{fig:basel}
	\end{minipage}
	\hfill
  \begin{minipage}[t]{0.45\linewidth}
	   \includegraphics[width=0.7\linewidth]{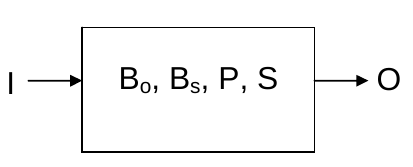}
	   \caption{Representation of a time-discrete and -invariant basic functional element with state and parameters}
	   \label{fig:basel1}
	\end{minipage}
\end{figure}

Embedded systems can be modelled by a \textit{network of basic functional elements}, like shown in Figure \ref{fig:controller} for a simple closed control loop. Adapted from \cite{Willems1991}, a basic functional element can be generally represented by a tuple $\langle B, I, O \rangle$, where $B$ represents the behavior set, $I$ the input set and $O$ the output set, according to the following relationship (Figure \ref{fig:basel}):
\[
o(t) = b\Big(t, i(0..t)\Big) \text{ with }o \in O, i \in I, b \in B, t: \text{time} 
\]

This representation is also closely related to the definition of an actor model as can be found e.g. in \cite{Lee2017} (see Section \ref{ssec:ModelEmbed} for details).

If we assume time-discrete systems (which is usually true for embedded system control) and time-invariant functional behavior (which means the same input history $i(0..t)$ produces always the same output $o(t)$), we can subsume the input history $i(0..t)$ to a state $s(t)$. Furthermore, a time invariant parameter set $P$ can be often split from the input set $I$ (e.g. the values for P, I and D in the above controller example). This results in the representation of a basic functional element by the tuple $\langle B_o, B_s, S, I, P, O \rangle$, where $B_o$ represents the output behavior set, $B_s$ the state behavior set, $S$ the state set, $I$ the input set, $P$ the parameter set and $O$ the output set (Figure \ref{fig:basel1}):
\begin{align}
o(t) &= b_o\Big(s(t), i(t), p\Big) \notag \\
s(t+1) &= b_s\Big(s(t), i(t), p\Big) \text{ with }o \in O, i \in I, s \in S, p \in P, b_o \in B_o, b_s \in B_s, t: \text{timestep} \notag
\end{align}
 
\begin{figure}[ht]
     \centering
	   \includegraphics[width=0.5\linewidth]{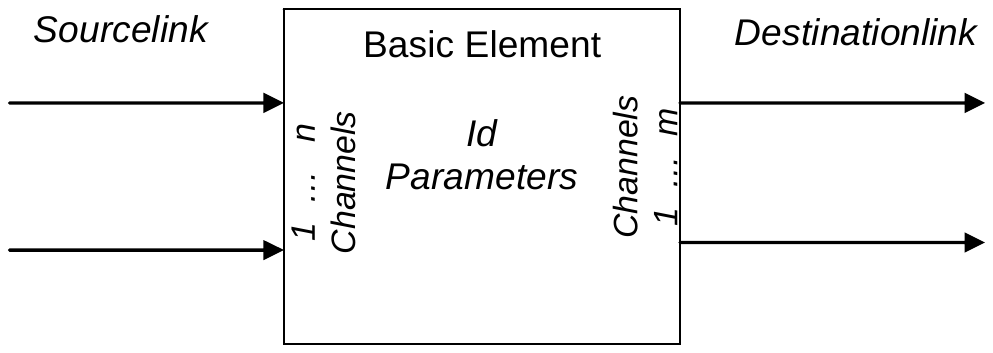}
	   \caption{Structure of a basic element for system description}
	   \label{fig:basicelement}
\end{figure}

The approach presented here is based on the observation that only a limited amount of different basic functional element behaviors is necessary to compose a wide range of embedded systems. This is a well known concept in embedded systems design. To simplify wording, from now on we call a basic functional element with a specific behavior simply a \textit{basic element}. Examples for such basic elements would be filters, controllers, arithmetic/logic units, sensors, actors, etc. Due to the limited number of different behaviors, a specific basic element can be identified by a \textit{unique id}, which defines its behavior (e.g. 1 = arithmetic/logic unit, 10 = PID controller, 500 = sensor, 600 = actor, ...). Additionally, the corresponding \textit{parameter set} (e.g. the arithmetic logic operation of an arithmetic/logic unit, the values of P, I, D and the loop period of a PID controller, the resource\footnote{the physical address of a specific sensor, e.g. engine temperature sensor 5.} and sample period of a sensor, ...) has to be given. Not all basic elements will have a state (e.g. an arithmetic/logic unit is usually a pure combinatorial unit with no state). Basic elements with state will start with an initial state-value (e.g. 0 for the integral sum of a PID controller), which might be influenced by the parameters. Figure \ref{fig:basicelement} shows such a basic element 
with n \textit{source link channels} representing the input $I$, m \textit{destination link channels} representing the output $O$, the \textit{unique Id} to represent its behavior ($B_o, B_s$) and the \textit{parameters} representing its parameter set $P$. The state $S$ is an implicit value. 

\begin{figure}[ht]
     \centering
	   \includegraphics[width=0.4\linewidth]{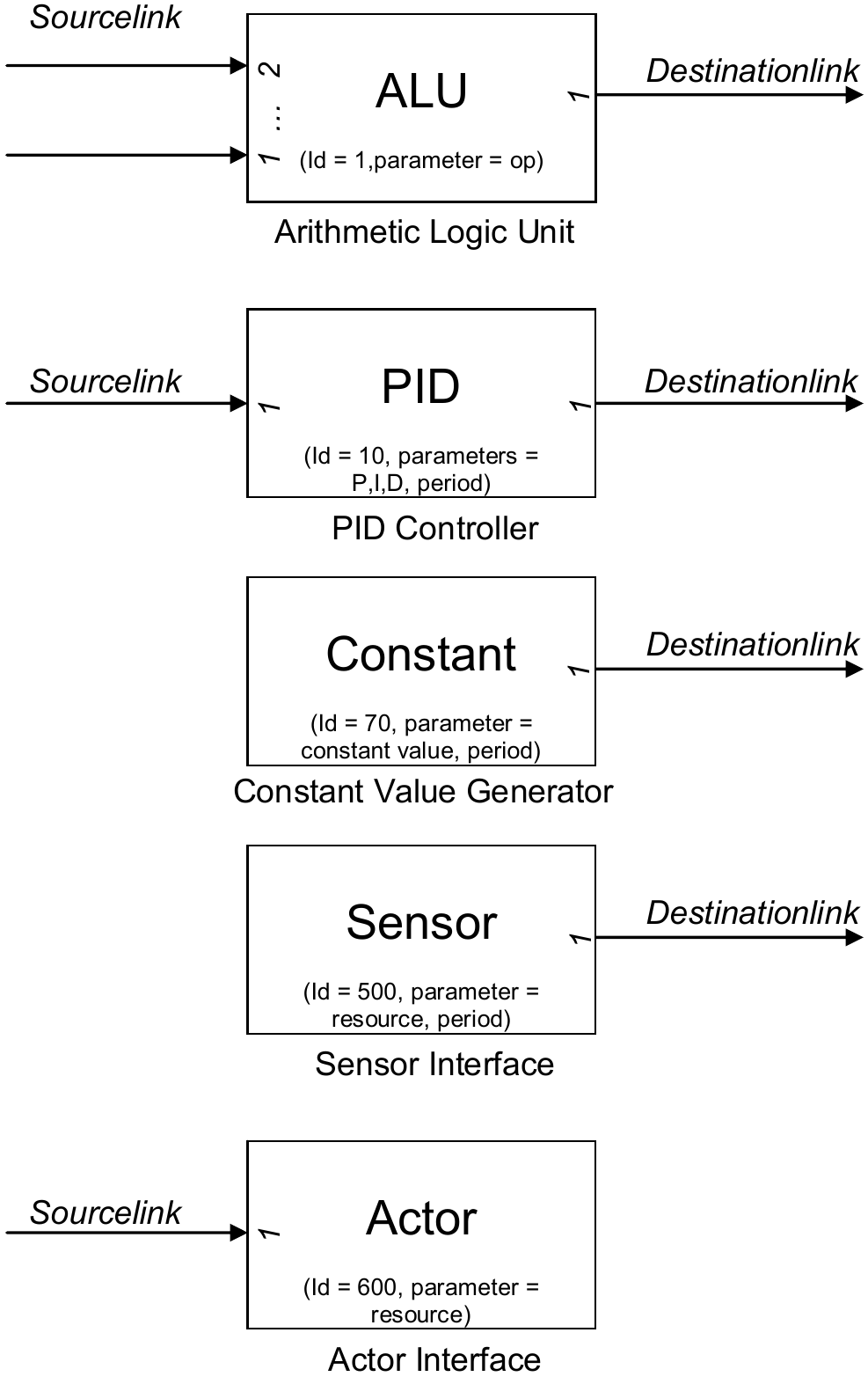}
	   \caption{Sample basic elements}
	   \label{fig:samples1}
\end{figure}

Figure \ref{fig:samples1} shows examples, some of them were already mentioned above. The Id numbers are arbitrarily chosen, it is only important that they are unique. We have an arithmetic logic unit (ALU) element with the Id = 1 and the parameter defining the requested operation (minus, plus, mult, div, greater, ...). The two input operands are given by channels 1 and 2 of the sourcelink whereas the result is provided via a single-channel destinationlink. Such an element is needed for calculations in a dataflow, e.g. a setpoint comparison in a closed control loop. Another element often used in closed control loops is the PID controller. Here, this element has the unique Id = 10 and the parameter values for P, I, D and the control period. Furthermore, it has a single channel source- and destinationlink. Other popular elements in embedded systems are interfaces to sensors (Id = 500, the parameters resource and period define the specific sensor and its sample period) and actors (Id = 600, resource specifies the specific actor) or a constant value generator (Id = 70, the parameters are the output value produced and its period).

\subsection{Basic Idea}
\label{sec:basicidea}

If a sufficient set of these basic elements is provided, a wide range of embedded systems can be completely composed by simply combining and parametrizing these elements. The basic elements are used as pre-defined building blocks. This fact is also exploited in model driven design, e.g. by constructing a Matlab model. However, in our approach we use it to generate a compact description of the targeted embedded system which can be stored in each processor core to serve as an digital artificial DNA. This enables the \textit{self-building of the system at run-time} which provides several advantages compared to classical design approaches:
\begin{itemize}
\item Since the system builds itself at run-time, it can \textit{adapt at starting time to the current state of the hardware platform}. Each processor core knows the entire artificial DNA, so it can calculate its own suitability for each basic element and compare it to the other cores. This can be done with respect to various criterias like computational power (clock frequency, available memory, floating point processing, energy consumption, ...), actual state (load, temperature, available energy, ...) and communication links (directly interacting basic elements should be placed close to each other). Therefore, the systems adapts itself autonomously to the available hardware platform (self-configuration). Furthermore, the system can \textit{react to changes in the hardware platform at runtime}. If e.g. a new processor core is added, it can copy the current artificial DNA from any other processor core in the system thus integrating itself into the system. As we show later, artificial DNAs are rather small (usually $< 1kByte)$, so the copy process is simple. If the state of processor cores change at run-time (e.g. temperature increase, energy shortage, ...), the system can react autonomously by recalculating the suitabilty of cores and relocating basic elements (self-optimization). If cores fail or are removed, the system can reshape itself to the remaining cores according to its artificial DNA (self-healing). The system can also react to changes in the communication links. \textit{This results in a very robust and flexible system}. A classical design (e.g. a MatLab model with code generation and design time allocation) can't do that.
\item \textit{Setting up a system from a set of pre-defined basic elements} has a big advantage over combining individiual user-designed application specific elements (user-programmed tasks). For each pre-defined basic element in the set, the base suitability and the suitability/state relationship can be defined once for each available processor core type. 
Then, each processor core is able to calculate its own suitability automatically at run-time for each basic element (like mentioned above) in all applications composed from this set.
 In contrast, for individual user-designed application specific elements this has to be done individually by the user in each application\footnote{When using e.g. the AHS without the artificial DNA, the initial hormone values for each individual task have to be defined by the user. With the artificial DNA, these hormone values are calculated automatically for the basic elements.}. This gets more and more unfeasible when embedded systems are growing larger. 
\item The artificial DNA is a \textit{compact} (see Sections \ref{ssec:artDNA} and \ref{sec:moreDNAs}) and \textit{platform independent} representation of an embedded system, which connects and parameterizes simple basic elements (representing the vocabulary of a common language) at runtime to form an application. Arbitrary heterogenous platforms are supported, each processor just has to provide the common set of basic elements (the vocabulary), usually in form of a library\footnote{This library also contains the processor/core type base suitability and the suitability/state relationship for each basic element, as mentioned in the previous bullet item.}. This library is also rather small ($30 - 70 kBytes$), as prototypic implementations have shown (see Section \ref{sec:implementation} and \cite{BrinksJournal2017}). The separation between basic elements and their connection and parameterization also simplifies development and testing. Basic elements can be tested independently from a concrete artificial DNA. The same basic elements are reused for a wide range of different artificial DNAs. Depending on the field of application, different basic element libraries can be created. Using the concept of artificial DNA, the building plan of an application can be modified at run-time (by changing the connection or parameterization of the basic elements) or new versions of the basic elements themselves can be applied.
\item The artificial DNA is a \textit{fine grain representation of the system functionalities} using simple basic elements. This offers flexibilty in assigning such functionalities to processor cores. A functionality (like e.g. an ABS brake functionality for cars) not necessarily needs to be assigned to a single processor core, it can be \textit{spread among the available processor cores at the granularity level of basic elements at run-time} in the best possible way. This is different to classical approaches, where e.g. in the automotive area a functionality like ABS is bound to a fixed ECU with a redundant backup ECU. In case of failure, the complete functionality has to be taken over by the backup ECU, if both ECU and backup ECU fail the functionality is lost.  Using the artificial DNA, all system functionalities share all cores at basic element granularity level. In case of a core failure only the affected basic elements have to be taken over by other cores. All functionalities stay operational as long as enough cores are available, there are no specific cores for specific functionalities. This keeps the system running as long as possible. The fine grained representation based on simple basic elements also allows to \textit{exploit many-core architectures} in an efficient way.
\item The artificial DNA determines the functionalities of the embedded system. By loading a new or modifying the current artificial DNA, the \textit{functionalities can be modified at run-time}. This simply enables \textit{dynamic reconfiguration}. Furthermore, it enables \textit{system evolution} at run-time. Genetic algorithms can be used to evolve an embedded system under operation. This is a interesting and challenging field of research, especially with respect to very large and complex embedded systems. It will be a major focus of our future work (see section \ref{sec:futurework}).
\item Besides the already mentioned self-healing to compensate core-failures, more \textit{fault-tolerance mechansims can be synthesized at run-time} using the artificial DNA. Assume the system designer defines an importance value of a system output (usually an actor in the system, e.g. a braking actor of a car). Then the processor cores are able to backtrack this importance value along the artificial DNA back to the inputs and to assign it to each affected basic element. Now, depending on importance values and available resources (number and state of the available processor cores) very important basic elements in the current artificial DNA can automatically be replicated at run-time (for double or triple modular redundancy) to identify and remove errors and error sources (e.g. unreliable processor cores). This opens a wide range of graceful degradation: as long as enough processor cores are available, all functionalities are built and important functionalities are protected by redundancy. As the number of cores is reduced by failures, first the redundancy is reduced while all functionalities are maintained. If the number of available cores falls below a level where all functionalities can be maintained, the less important functionalities are dropped first.
Another possibility of extended fault tolerance is using the implicit redundancy given by an artificial DNA. Since an artificial DNA consists of rather simple basic elements, most of these elements will occur multiple times in an artificial DNA (e.g. an arithmetic/logic unit, a PID controller, ...). These multiple occuring elements could check themselves mutually, e.g. by exchanging challences when they are not busy. Such extended self-healing mechanisms are another interesting research field, which will be addressed by our future work (see section \ref{sec:futurework}).
\end{itemize}

\subsection{Artificial DNA}
\label{ssec:artDNA}

\begin{figure}[ht]
     \centering
	   \includegraphics[width=0.7\linewidth]{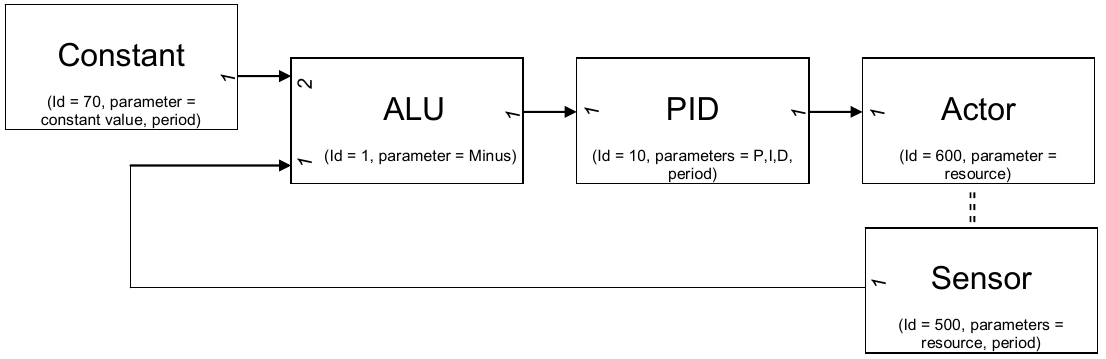}
	   \caption{The closed control loop consisting of basic elements}
	   \label{fig:simpleexample}
\end{figure}

\begin{figure}[ht]
     \centering
	   \includegraphics[width=0.7\linewidth]{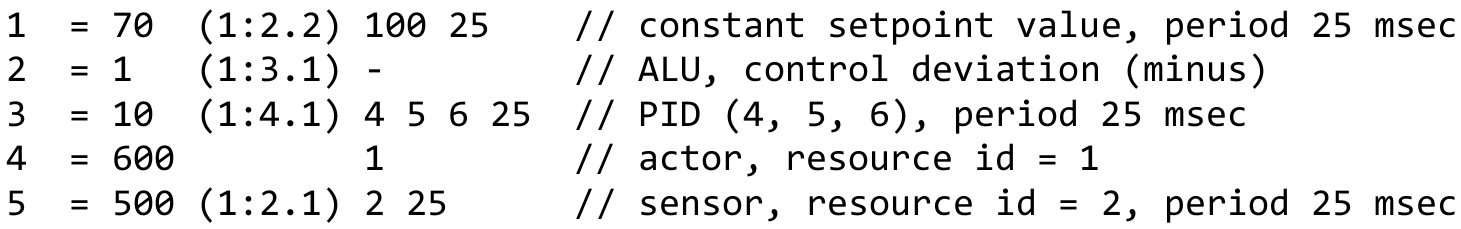}
	   \caption{Artificial DNA representation of the closed control loop}
	   \label{fig:sampledna}
\end{figure}

The building plan of an embedded system can be described by an extended netlist, which contains the interconnections and the parameters of the basic elements used. This extended netlist represents the digital artificial DNA of the system, since it can be used to completely build up the system at run-time. Even for complex systems, this artificial DNA is small enough ($< 1 kByte$, see Section \ref{sec:moreDNAs} and \cite{BrinksJournal2017}) to be stored in each processor core like the biological DNA is stored in each cell. In this way the embedded system becomes \textit{self-describing}. Each line of the artificial DNA contains the Id of a basic element, its connection to other basic elements (by defining the corresponding destinationlinks for each sourcelink of the basic element) and its parameters:
\begin{small}
\\
\\
\textit{Artificial DNA line = Id Destinationlink Parameters}
\\
\\
\end{small}
The destinationlink description in an artificial DNA line can be defined as the following set:
\begin{small}
\\
\\
\textit{(Destinationlinkchannel:Destination.Sourcelinkchannel ...)}
\\
\\
\end{small}
Here, \textit{Destinationlinkchannel} gives the channel number of the destinationlink, \textit{Destination} refers to the line of the basic element the destinationlink channel is connected to and \textit{Sourcelinkchannel} is the channel number of the sourcelink channel of the destination element. As an example, the destinationlink description \textit{(1:10.1 1:9.2 2:7.1)} defines that channel $1$ of the destinationlink is connected to the sourcelink channel $1$ of the basic element in line $10$ of the DNA (\textit{1:10.1}) and to the sourcelink channel $2$ of the basic element in line $9$ (\textit{1:9.2}) while channel $2$ of the destinationlink is connected to the sourcelink channel $1$ of the basic element in line $7$ (\textit{2:7.1}). 
As an example, figure \ref{fig:simpleexample} shows the building plan of the simple closed control loop from figure \ref{fig:controller} using basic elements as building blocks. Now, Figure \ref{fig:sampledna} shows the artificial DNA of the system enriched with comments (defined by //). More examples can be found in Section \ref{sec:moreDNAs}, \cite{brinksCandC} and \cite{BrinksJournal2017}.

Even in case of a very special embedded system not being able to be composed from the set of standardized basic elements, special Ids for user/application specific elements can be defined to solve this issue.

\begin{figure}[ht]
	   \centering
	   \includegraphics[width=0.5\linewidth]{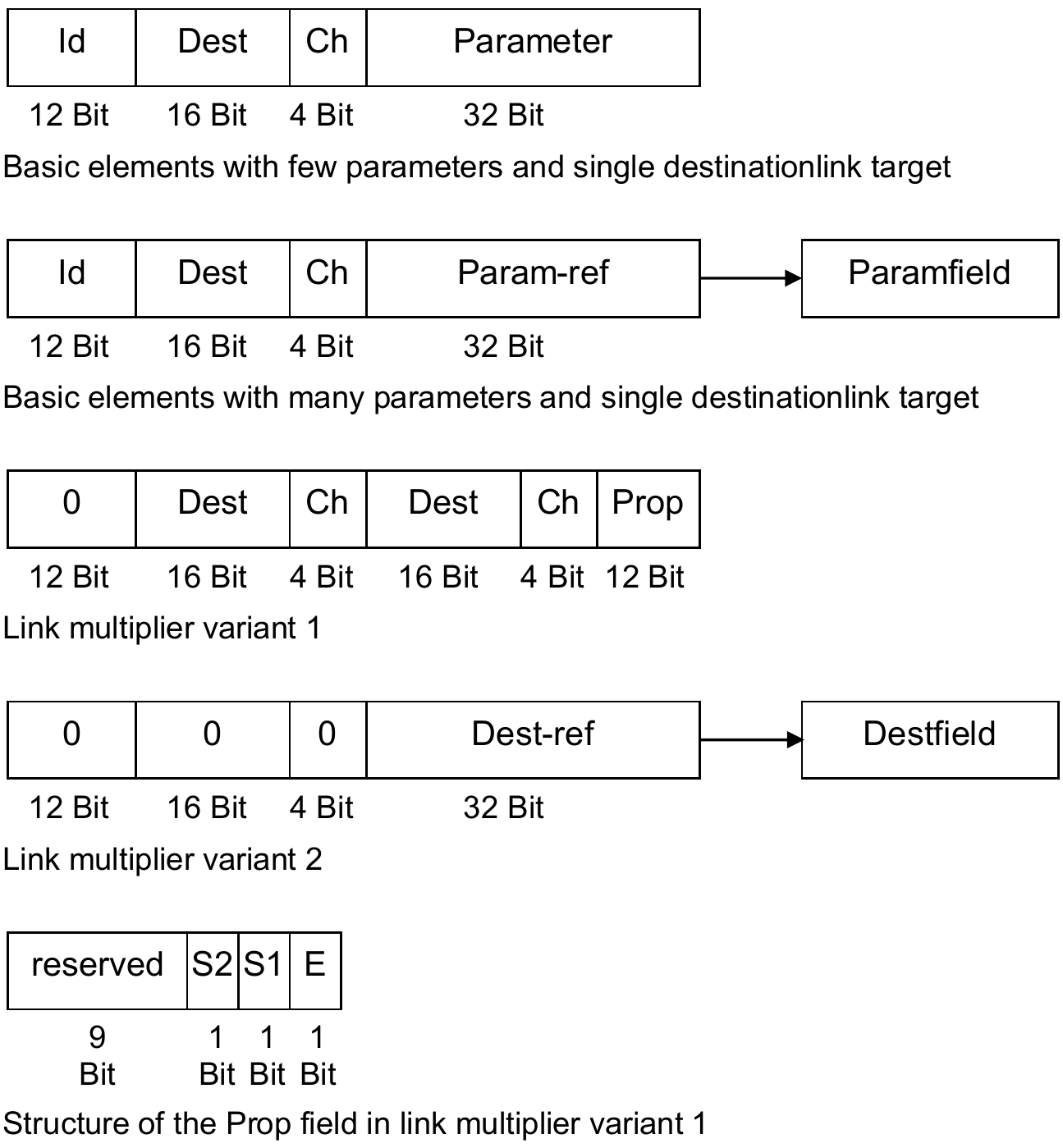}
	   \caption{Compact processor core internal storage format of the artificial DNA}
	   \label{fig:implementation}
\end{figure}

Inside a processsor core, the text-based description of the artificial DNA from Figure \ref{fig:sampledna} can be stored even more memory-efficient. Figure \ref{fig:implementation} shows a proposal for such an efficient and compact implementation of the artificial DNA. The first row of the picture sketches the primal implementation of a DNA line. A 12 Bit Id field initially allows to have up to 4096 different basic elements. This should be far enough to build embedded systems. The 16 Bit destination field (Dest) specifies the target DNA linenumber of the destinationlink channel 1 for this basic element thus allowing up to 65536 basic elements in the system. The 4 Bit channel field (Ch) specifies the sourcelink channel number of the target element. The last 32 Bit field specifies the parameters. Therefore, a DNA line can be coded in 64 Bits. If more than 32 Bit parameter space is required, a 32 Bit pointer to a separate parameter field can be specified instead of directly having the parameters in the DNA line. This is shown in the second row of Figure \ref{fig:implementation}. It keeps constant the size of a DNA line at 64 Bits allowing easy processing and browsing of the DNA. The Id field indicates whether the parameters of a DNA line are direct or indirect. This is easy to achieve since it only depends on the Id of a basic element how many and what type of parameters are needed. Finally, we have to cover the case of having more than one destinationlink target (multiple destinationlink channels, connection to multiple sourcelinks). This can be realized by a link multiplier as shown in the third row of Figure \ref{fig:implementation}. The destinationlink field of a basic element needing more than a single destinationlink target points to a DNA line containing a link multiplier. A link multiplier is identified by the special Id 0 and contains two destinationlink fields. This is sufficient for many cases. If more than two targets are needed, the link multiplier can be extended to the next line. This is defined by the $E$ Bit (\textit{E}xtend) in the Prop field of the multiplier as shown in the fifth row of Figure \ref{fig:implementation}. Setting this Bit indicates the multipiler is extended to the next line\footnote{Another possibility for extension is to use the second destination field of the multiplier as a pointer to another multiplier. This is less efficient since then all but the last multipliers can handle a single entry only. However, they now could be placed in a non consecutive order in the DNA. This might be advantegous when changing the DNA at run-time comes into play.}. So any number of destinationlink targets can be realized. The $S1$ Bit (\textit{S}eparate\textit{1}) in the Prop field defines if the first and second destinationlink entry in the multiplier refer to the same or different destinationlink channels. If the Bit is not set, they belong to the same channel. In case of extending the multiplier to the next line, the $S2$ Bit (\textit{S}eparate\textit{2}) defines if the first destinationlink entry in the next line refers to the same or different destinationlink channels. More information and an example of this internal representation can be found in \cite{brinksCandC}.

\subsection{Building the System from its Artificial DNA}
\label{sec:build}

\begin{figure}[ht]
     \centering
	   \includegraphics[width=0.5\linewidth]{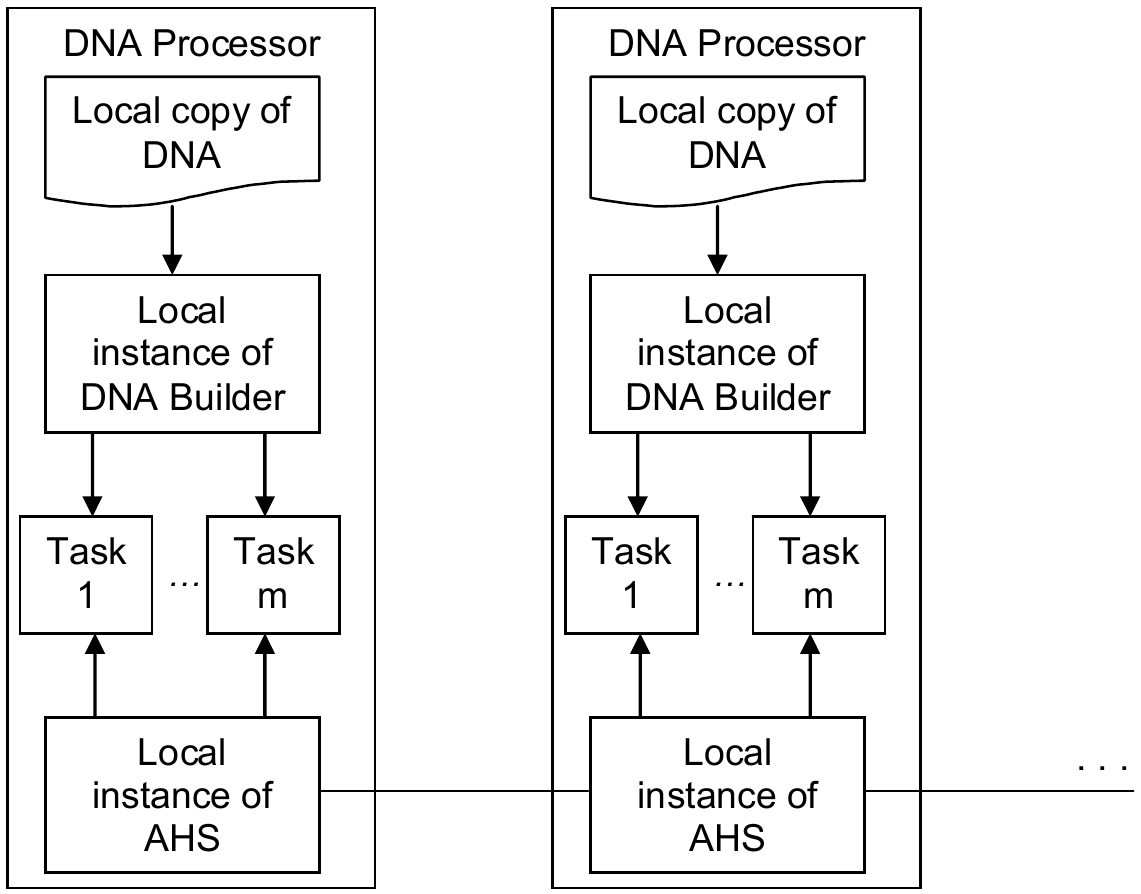}
	   \caption{System architecture}
	   \label{fig:architecture}
\end{figure}

Using the artificial DNA the system now becomes \textit{self-building} at run-time. The DNA serves as the basis for the middleware layer of the distributed embedded system to setup and connect the system tasks. Figure \ref{fig:architecture} shows the system architecture using a DNA builder and the AHS as middleware layer. There, we call the processor cores of the distributed system DNA processors. 

First, the DNA builder parses the DNA and segments the system into concurrent tasks. 
The artificial DNA is a functional description of the system including all dependencies, so it is a suitable guideline for task segmentation. The simplest possibility is to make each instance of a basic element a task in the embedded system. This is how the current prototypic implementation works. However, for future implementations it would be also possible to apply a larger granularity and combine multiple directly interacting basic elements into a single task to reduce the number of concurrent tasks and task communication overhead.

Second, the AHS tries to assign tasks to the most suitable processor cores. With the artificial DNA, the suitability of a processor core for a basic element and therefore the corresponding task can be derived automatically (see also previous section) by the DNA builder from the Id of the basic element and the features and state of the processor core. As an example, a basic element with Id = 10 (PID controller) performs better on a processor core with better arithmetic features while memory is less important. For the AHS as underlying middleware, core suitability is indicated by specific hormone levels \cite{UB-SPRINGER-2008}. So the appropriate hormone levels can be calculated automatically by the DNA builder and assigned to the AHS. 

Third, task relationship is also considered for task assignment. The AHS tries to locate cooperating tasks in the neighborhood to minimize communication distances. This has to be indicated also by hormone levels \cite{UB-SPRINGER-2008}. Using the artificial DNA, task relationship can be derived automatically by the DNA builder from analyzing the destinationlink fields of the DNA lines. This allows to setup the communication links between tasks and to determine cooperating tasks. So the appropriate hormone levels can be generated automatically. 

All steps of this building process are linear in time with relation to the number of basic elements $n$, so the overall time complexity is $\mathcal O(n)$. 

In case of changes in the hardware platform at runtime (e.g. loosing or gaining DNA processors\footnote{The loss of DNA processors (e.g. by a permanent processor failure) is detected by missing hormones of the AHS, newly arriving cores are detected by additional hormones of the AHS.}) the system autonomously restores or readapts itself by the DNA which is present in each DNA processor. The time complexity for this is also $\mathcal O(n)$, here $n$ is the number of affected basic elements (e.g. the number of basic elements lost by a failed DNA processor).

Overall, the artificial DNA represents the blueprint that enables the self-building of an embedded system application at run-time. This enables a large amount of adaptivity, flexibility, dependability and robustness and is a leverage to handle large distributed embedded systems.

\section{Related Work}
\label{sec:relwork}
The concept presented in this paper covers topics in the fields of Organic Computing, DNA Computing, modelling of embedded systems and fault tolerant computing. In the following, related work and the differences to this approach are discussed.

\subsection{Self-Organization and Organic Computing}
\label{ssec:OrganicComp}

\textit{Self-organization} has been a research focus for several years. Publications like \cite{JE-DEUTSCH-1989} or \cite{WI-ACM-1995} deal with 
basic principles of self-organizing systems, like e.g.\ emergent behavior, reproduction, etc. Regarding 
self-organization in computer science, several projects and initiatives can be listed. 

IBM's and DARPAS's \textit{Autonomic Computing} project \cite{IBM-AUTONOMIC}, \cite{JK-Autonomic-2003} deals with self-organization of IT servers in networks. Several so-called \textit{self-X properties} 
like self-optimization, self-stabilization, self-configuration, self-protection and self-healing have been postulated.
The \textit{MAPE cycle} consisting of \textit{M}onitor, \textit{A}nalyze, \textit{P}lan and \textit{E}xecute was defined to realize these properties. It is executed in the background and in parallel to normal server activities similar to the autonomic nervous system. 

The German \textit{Organic Computing} Initiative was founded in 2003. Its basic aim is to improve the controllability 
of complex embedded systems by using principles found in organic entities \cite{VDE-OC-2003,SCH-ISORC-2005}. Organization principles which are successful in biology 
are adapted to embedded computing systems. The DFG priority programme 1183 "Organic Computing" 
\cite{DFG-SPP-1183} has been established to deepen research on this topic.

Self-organization aspects for computing systems are
also addressed by several international research programs, e.g.\ \cite{EU-FET,CS-COMPLEX}.

\textit{Self-organization for embedded systems} has been addressed especially at the ESOS workshop \cite{ESOS-2013}. Furthermore, there are several projects related to this topic like ASOC \cite{HE-AC-2005,HE-SASO-2009}, CARSoC \cite{TU-ACACES-2006,TU-SEUS-2008} or DoDOrg \cite{UB-ARCS-2006}. In the frame of the DoDOrg project, the \textit{Artifical Hormone System} AHS was introduced \cite{UB-ARCS-2006,UB-SPRINGER-2008}. Another hormone based approach has been proposed in \cite{Trum2006}. \cite{WEISS-ZELLER-2009} describes self-organization in automotive embedded system. 

Another ongoing research project using organic computing principles is \textit{Invasic} \cite{DFG-SPP-89}. This DFG funded transregional investigates mechanisms how future parallel computing systems can be exploited by resource aware programming. Applications invade, infect and finaly retreat from parallel computing resources. This project also shows the high importance of organic computing principles for future highly parallel and distributed systems.

Different to our approach, none of the approaches described above deal with self-description or self-building using DNA-like structures. 

\subsection{DNA Computing}
\label{ssec:DNAComp}

\textit{DNA Computing} \cite{dnaGarzon} uses molecular biology instead of silicon based chips for computation purposes. Theoretical and application oriented approaches can be distinguished. Theoretical approaches investigate how information can be coded by DNA or special languages and grammers are discussed. Application oriented approaches try to solve e.g. optimization problems using molecular biology. In \cite{Lee200439}, the traveling salesman problem is solved by DNA molecules. Our approach relies on classical computing hardware using DNA-like structures for the description and building of the system. This enhances the self-organization and self-healing features of embedded systems, especially when these systems are getting more and more complex and difficult to handle using conventional techniques. 

Our approach is also different from \textit{generative descriptions} \cite{hornby2001}, where production rules are used to produce different arbitrary entities (e.g. robots) while we are using DNA as a building plan for a dedicated embedded system.

\textit{Architectural models} \cite{Shaw1996} describe system components and their behavior on an abstract level. A major problem of these models is to prove the correspondence of the abstract architecture with the concrete system. A solution is to transform the architectural models to language constructs, see e.g. \cite{archjava1}, \cite{archjava2}. Another possible solution is to monitor the system at run-time and to draw conclusions on architectural properties \cite{garlan2004}. These approaches are different from the approach presented here. The artificial DNA is focused on the domain of embedded systems which enables well defined design patterns and templates. Therefore the artificial DNA can represent an exact building plan of the embedded target system. The focus of architectural models is much broader (software), therefore the validation of these models is very complex.

\subsection{Design and Modelling of Embedded Systems}
\label{ssec:ModelEmbed}
There are several well researched techniques to model and design embedded system. The \textit{model-based design} approach \cite{Model2010} describes the embededed system on a higher abstraction level using a domain-specific language (like e.g. Matlab/Simulink). Based on this description the embedded system is gradually refined and compiled to the target platform, which might be software for specific processors or hardware description (e.g. VHDL) for an FPGA/ASIC synthesis (hardware/software codesign). This allows to adapt the mapping of software and hardware parts at compile time.

The \textit{platform-based design} approach \cite{Vincentelli2001} uses abstraction layers (the platforms) to constrain design choices during the refinement process in the design flow. By defining the mapping between these layers the system is refined and compiled from the application layer to the architectural layer. According to \cite{Lee2003}, model-based and platform-based design can be seen as two sides of the same coin, where platform-based design puts its main focus on physically realizable platforms while model-based design puts its main focus on the application space.

\textit{Actor-oriented design} \cite{Lee2003} is a component methodology close to the application space. Actors are inspired by physical models of an embedded systems (see e.g. \cite{Lee2017}), where system components are descriped as functions with input, output, state and parameters. Actor-oriented design separates the actor definition from the actor composition. Different computational models can be used when compiling actor definitions and an actor composition to the target system. One possibility is the use of the \textit{actor model}\footnote{The term 'actor' in actor-oriented design has been chosen from the actor definition in the actor model.} originally proposed for Artificial Inteligence \cite{Hewitt1973}. Here, actors are concurrent processes communicating asynchrounosly via messages in mailboxes. Messages can be received out-of-order. Actors can send messages, react to messages, create new actors and change their own behavior. \textit{Kahn process networks} \cite{Kahn74} are a more restrictive computational model. Here, processes (actors) communicate via unbounded FIFOs, so message order is preserved. While write operations never block, read operations always stall when no message is present. While Kahn process networks are completely deterministic, they are difficult to schedule on real platforms. \textit{Dataflow networks} \cite{Lee1995} are an easier to schedule variant. Here, a process (actor) fires (starts to operate) when all input data is available. In \textit{synchronous dataflow networks}, additionally the amount of data produced and consumed is fixed by a contract allowing static scheduling of processes and memory resources. This is a commonly used computational model for actor-oriented design. \textit{Discrete event models} are using timestamped events (including values) for communication between actors. This is mostly used when translating actors to hardware (VHDL), where the actors internally operate synchronously with their own clock while the clocks between the actors are asynchronous. \textit{Synchronous models} assume that all actors operate synchronously by a common clock.
\\
\\
The approach presented here is different in the following aspects:

The basic idea of our approach is to mimic the biological DNA by storing the complete building plan of an embedded system in each processor core. Based on this building plan, the \textit{system builds itself dynamically at run-time in a self-organizing way} in adaption to the currently available hardware platform. This is a major difference to techniques described above, where the mapping of the desired system to the hardware platform is done by tools at design time (e.g. a Matlab/Simulink model). The building plan acts as artificial DNA. It shapes the system autonomously to the available distributed multi/many-core hardware platform and re-shapes it in case of platform and environment changes (e.g. core failures, temperature hotspots, energy shortage, reconfigurations like adding new cores, removing cores, changing core connections. etc.). Our approach provides self-configuration, self-optimization and self-healing at run-time while still maintaining real-time capabilities.

To realize the artificial DNA, we have to describe the building plan of an embedded system in a compact way so it can be stored in each processor core. Therefore, we have chosen to represent the artificial DNA as a \textit{simple netlist} \cite{netzliste} of basic elements defined by unique numerical identifiers and their parameters. The linear structure of this netlist it also more suitable for run-time DNA modifications like dynamic reconfiguration or evolutionary mechanisms\footnote{Such linear structures can be easily handled by e.g. the coding of evolutionary algotrithms or learning classifier systems.} as the approach presented in \cite{Lee2003}, where the composition of elements is described in a nested XML scheme or a programming language.

Since actors as presented in \cite{Lee2017} are very convenient components to compose embedded systems, we have adopted this concept to form our basic elements. However, in contrast to e.g. \cite{Lee2003}, the basic elements in our approach are \textit{simple and generic paramterizable executable modules} residing in a small run-time library. In combination with the composition by a linear netlist, this is the key enabler for building, adaption, reconfiguration and evolution of the system at run-time.

The computational model of our approach is also slightly different from the ones described above. It is a \textit{combination of a dataflow-driven and time-driven network}\footnote{A time-driven network can be considered as a special case of an discrete-event network, where the events are restricted to purely timed events generated by an internal clock of the basic element.}. To clearify this, here are some examples from basic elements described in Section \ref{sec:compmod}: the ALU is a pure dataflow-driven basic element, a new new result is sent as soon as new operands are available. The Sensor is a pure time-driven basic element, a new value is sent each time-period defined by a parameter. The PID controller can be used both ways, as defined by the parameters. In the time-driven mode, the PID controller delivers an output each time-period using the latest available input value. This ensures a constant period which is important for a closed contol loop. In the dataflow-driven mode, it delivers a new output as soon as a new input is received assuming the period is ensured by the input data (e.g. from a time-driven element like a sensor). Another example for a mixed mode element is the Counter basic element. In the time-driven mode it delivers timed output counts, while in the dataflow-driven mode it counts arriving inputs. This combined computation model has shown to be very efficient and flexible regarding use and implementation of the basic elements in the prototypic implementation (see Section \ref{sec:implementation}).

Overall, our approach combines Organic Computing principles with the field of Model- respectively Actor-oriented design to gain the advantages from both areas. 

\subsection{Fault Tolerant Computing}
\label{ssec:FaultTolerant}
Failure detection is an important basis for fault tolerant computing. Classical approaches in failure detection reach from
simple parity checking to more advanced error detecting (and partly correcting) codes like CRC, Reed Solomon, etc.
These techniques are commonly used to detect failures in communication or memory systems. Error detecting and correcting
codes can be implemented in software or with hardware support. Approaches introducing special hardware
units to detect and correct errors are e.g.\ \cite{SS-VLSI-2007}, where such codes are used for failure handling
in on-chip busses, or \cite{AB-VEHI-2007} applying error detecting codes in LDPC (\textit{L}ow \textit{D}ensity \textit{P}arity \textit{C}heck) decoders.

Regarding more general failure detection techniques for hardware circuits, three basic principles can be distinguished: Firstly,
failure detection by spatial hardware redundancy (duplication of hardware circuits and result comparison)
has to be mentioned. Using this technique, permanent, transient and timing errors can be detected. A discussion
can be found e.g.\ in \cite{MM-TEST-2000}. A disadvantage of spatial redundancy is the 
considerable overhead in hardware and energy consumption. Secondly, failures can be detected by 
time redundancy. The use of shadow registers is such an approach to detect transient failures.
Here, values (e.g.\ in a processor pipeline) are stored in a shadow register with a cycle delay thus
delivering differences to the value in the original register in case of a transient failure.
Examples for this technique can be found in \cite{AN-DATE-2000}, \cite{EK-MICRO-2003} or \cite{AB-COMP-2004}.
Time redundancy usually introduces less overhead than spatial redundancy.
Third, so-called self-checking circuits can be used. This is often done for arithmetic or logic circuits,
where the results can be checked according to mathematical rules. This technique can be found e.g.\ in 
\cite{NN-DATE-1999,N-VLSI-2003,AN-VLSI-2000}.

Failure inhibition and correction can be done in several ways. An interesting approach inhibiting
transient failures in combinatorial logic by electrical and latching-window masking can be found
in \cite{SK-DEPEND-2002}. Here, failure models are used to inhibit transient failures 
for SRAM, latches and combinatorial logic. There are several other projects aiming to build
highly dependable systems based on failure correction. In \cite{TT-TRANS-1990}, failure correction
was done by rolling back operations in a CPU pipeline. As a major drawback, the rollback 
mechanism in the pipeline is time- and resource-consuming. A solution introducing less overhead
has been proposed in \cite{A-MICRO-1999}, where a dynamic verification based on an extended commit 
phase in the processor pipeline is used to allow only valid results to pass the pipeline.
A project conducted in the frame of the Organic Computing research program is the
"Autonomic System on Chip" (ASoC) project \cite{HE-AC-2005}. In this project, an autonomic layer
is put on top of the chip's functional layer. Autonomic elements in the autonomic layer are responsible
for monitoring and counteracting failures on the functional layer. A MAPE cycle known from IBM's Autonomic Computing is used
for that purpose. Furthermore, the autonomic elements are monitoring each other. As further failure 
handling techniques, fault tolerant CPU datapaths and time redundancy is used. Besides from pure
hardware architecture, the complete design and development cycle for reliable SoC is addressed in the ASoC project.
The "Embryonics and Immunotronics" project \cite{BO-EVOLV-2000} introduces antibodies inspired by the biological
immune system to detect faulty configurations in FPGAs. In \cite{PD-ASPLOS-2006}, a
defect tolerant SIMD architecture is proposed. A large number of limited capability nodes self-organize
to form SIMD operations, even in the presence of hardware failures.

Increasing reliability can also be addressed by software based approaches. An overview on techniques
regarding operating systems can be found in \cite{TH-COMP-2006}, while \cite{L-FOSE-2007} 
investigates the reliability aspect in software engineering. Handling transient failures
in real-time operating system is researched in several projects, e.g.\ \cite{IN-DATE-2006,IP-DATE-2006,KH-COMP-2003}.

General principles to design and engineer dependable systems can be found in \cite{B-SPRINGER-2004}.
The design of dependable embedded systems has been researched in the DFG priority program 1500 \cite{DFG-SPP-1500}. Several projects
have been conducted to investigate new fault tolerance principles for future nanoscale hardware architectures. The MixedCoreSoC project \cite{MixedCoreSoC2013} co-conducted by the author of this paper examines the use of an artifical hormone system for fault tolerance in mixed signal systems. However, no artificial DNA is used there. The Amrosia project \cite{Tahoori2013} exploits cross layer information to counteract ageing effects. The MIMODeS \cite{Wehn2012} project also investigates a cross layer aproach for the design of efficient, dependable VLSI architectures. A cross layer perspective to design reliable software for unreliable hardware is investigated in \cite{Henkel2016}. Operating system aspects to increase the dependability of embedded systeme are investigated in the ASTEROID \cite{Ernst2012} and the DanceOS \cite{Spin2017} projects. The list of all projects in the priority program can be found in \cite{DFG-SPP-1500}. 

Failure handling in our approach is different from the work presented above. We rely on bio-inspired self-organization 
based on the artificial DNA and the underlying artificial hormone system. Self-healing of permanent processor failures like core crashes is inherently detected by changing hormone levels and corrected by rebuilding or restructering the system
according to its artificial DNA. In future work we want to exploit the potential of the artifical DNA to automatically synthesize additional fault tolerant mechanisms at run-time according to the available system resources and the system state, see Section \ref{sec:basicidea}.

\section{Prototypic Implementation}
\label{sec:implementation}
As a proof of concept, we first have implemented a simulator for the DNA concept. The results have been published in \cite{brinksCandC}. This simulator was focused on the ability of self-building a system from its DNA and reconstructing it in case of component failures. So the basic elements were simply dummies in the DNA simulator which are allocated to processor cores, interconnected and visualized. They provided no real functionality. However, simulation results showed that these basic elements were properly allocated and interconnected by the DNA so self-building and self-repairing is possible. 

Encouraged by these promising results we have decided to implement a real prototype of the DNA concept. In this prototype the basic elements provide real functionality (e.g. an ALU, a PID controller, etc.) and interaction schemes, so working systems can emerge from a DNA. This allows for a far better evaluation than the simulator does. Communication and memory needs as well as real-time properties can be investigated on a real application example, see section \ref{sec:evaluation}.

\begin{figure}[ht!]
     \centering
	   \includegraphics[width=0.7\linewidth]{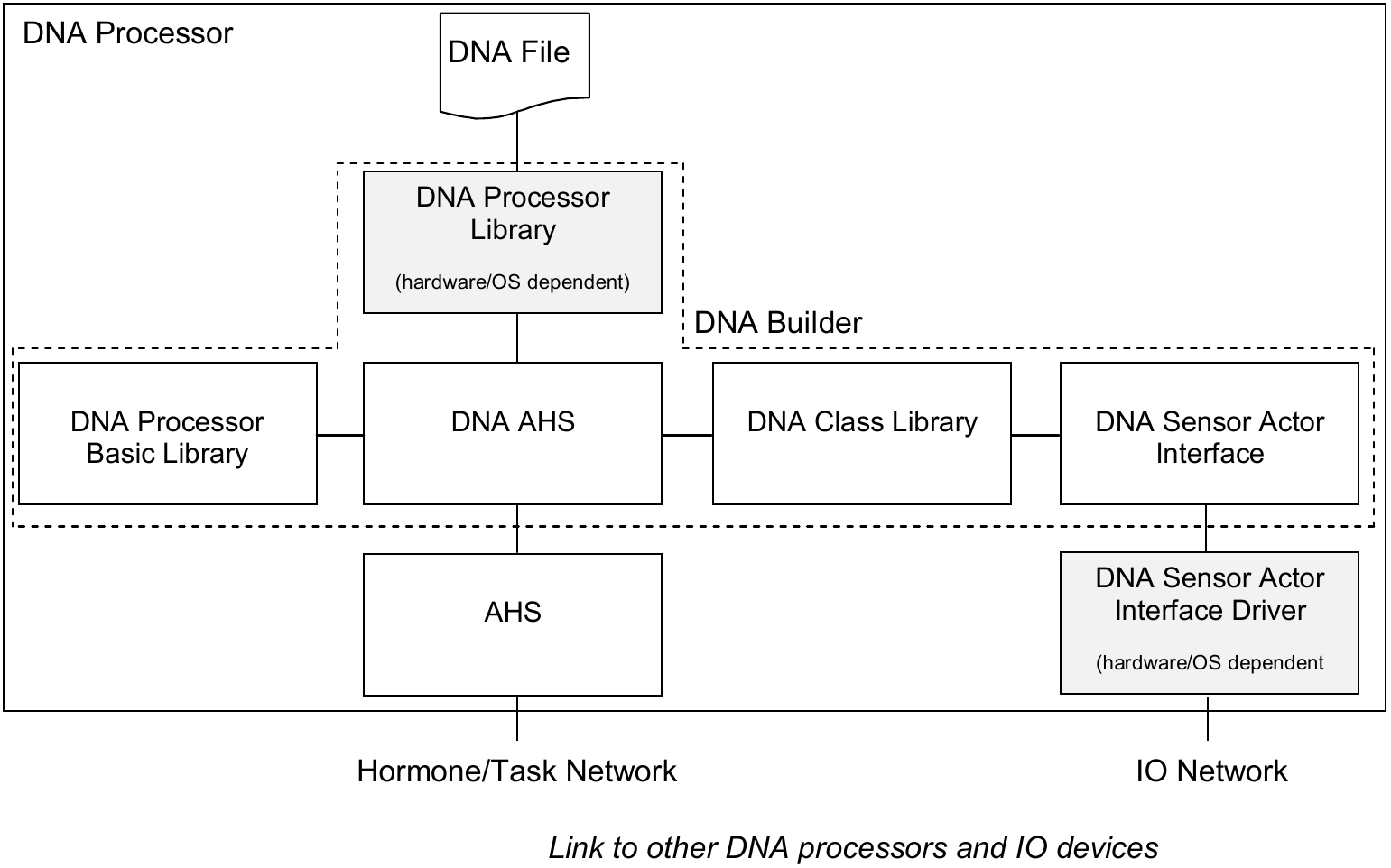}
	   \caption{Architecture of the DNA implementation}
	   \label{fig:DNAArch}
\end{figure}

Figure \ref{fig:DNAArch} shows the detailed architecture of a real DNA processor within the overall system architecture already presented in figure \ref{fig:architecture}. The currently active DNA is read from a file by the processor front end consisting of the \textit{DNA Processor Library} and the \textit{DNA Processor Basic Library}. While the first one contains all processor and OS specific parts, the latter is platform independent and provides generic processor-based functions like retrieving hormone values for basic elements on a given processor core (e.g. an ALU works better on a processor core with strong arithmetic features and therefore deserves higher hormone values to attract this basic element). This is done in cooperation with the \textit{DNA Class Library} which implements all the basic elements. Table \ref{table:classes} shows the basic elements realized in the prototypic implementation of the class library. In addition to the elements already mentioned in the previous section, there are elements to multiplex and demultiplex data values, to limit data values, to define thresholds, switching levels and hysteresis for data values. A complementary filter allows data fusion similar to but more simple than a Kalmann filter. The DNA checker creates a non-zero output value as soon as the system defined by the given DNA is completely set up and therefore becomes operational on the distributed DNA processors. It can be connected e.g. to an actor like a LED to indicate the operational state of the system contructed by the DNA (see figure \ref{fig:balancerDNA} in section \ref{sec:evaluation}). The DNA logger writes all input values to a log file and therefore allows the logging of data streams within the system. This small number of basic elements already enables a considerable range of embedded applications, as will be seen in section \ref{sec:evaluation}. All basic elements realized in this prototypic implementation use single precision IEEE float values for data exchange. 
The component \textit{DNA AHS} is the connector between all other components and the \textit{AHS Library}. Together these components realize the \textit{DNA Builder} introduced in section \ref{sec:build}. Based on the DNA read from file, all necessary basic elements are selected, all interconnections between these basic elements are defined and all hormone values are calculated. This information is promoted to the AHS library which places the basic elements to the DNA processors (see also section \ref{sec:build}). To provide an interface to sensors and actors, the \textit{DNA Sensor Actor Interface} component is used. It maps the resource id used as an abstract identification of a specific sensor or actor (see section \ref{sec:compmod}) to the real sensor or actor. This is done in a flexible way by a mapping table allowing the use of various sensors and actors. The \textit{DNA Sensor Actor Driver} component is used to access the real sensor/actor hardware\footnote{This driver allows not only access to real sensors and actors, but also to simulated ones. So the implemented DNA system is able to handle mixed environments consisting of real and simulated hardware.}. 
Only the two components shaded gray in figure \ref{fig:DNAArch} are platform dependent, all other components are independent from the used processor, sensor/actor hardware and OS platform. This allows a high portability of the DNA processor implementation. All components are implemented in ANSI C/C++. We compiled them for two target platforms: PC running Windows and Raspberry Pi running Linux. Table \ref{table:components} shows the memory footprint of both implementations. It can be seen that this footprint is rather small and compact. Only the DNA processor library component for Windows is big. This is due to the fact that Microsoft Foundation Classes (MFC) are used there to show processor information in Windows dialog boxes. The Raspberry Pi implementation uses console IO for this purpose and is therefore rather small. The overall DNA processor size on the Raspberry Pi is only 270 kBytes. It is possible to create a similar small footprint for Windows PC by using a console version instead of MFC.
Both implementations are fully compatible and can be used in a distributed heterogeneous environment.

\begin{table}[ht!]
\centering
\caption{Basic elements implemented}
\label{table:classes}
\begin{tabular}{c | c || c | c}
\hline\noalign{\smallskip}
Id  & Basic Element & Id  & Basic Element\\
\noalign{\smallskip}\hline\noalign{\smallskip}
1   & ALU           & 50  & Complementary Filter\\
    &               &     & \\
10  & PID           & 70  & Constant\\
11  & P             & 71  & Counter\\
12  & I             &     & \\
13  & D             & 500 & Sensor\\
    &               & 600 & Actor\\
40  & Multiplexer   &     & \\
41  & Demultiplexer & 997 & Stop\\
42  & Level         & 998 & DNA Checker\\
43  & Limit         & 999 & DNA Logger\\
44  & Hysteresis    &     & \\
45  & Threshold     &     & \\
\noalign{\smallskip}\hline
\end{tabular}
\end{table}
 
\begin{table}[ht!]
\centering
\caption{Components of the DNA implementation}
\label{table:components}
\begin{tabular}{l | c | c}
\hline\noalign{\smallskip}
                            & Raspberry Pi & Windows PC\\
\noalign{\smallskip}\hline\noalign{\smallskip}
DNA-AHS                     & 31 kBytes  & 71 kBytes\\
DNA Class Library           & 34 kBytes  & 75 kBytes\\
DNA Processor Basic Library & 7 kBytes   & 19 kBytes\\
DNA Processor Library       & 27 kBytes  & 3964 kBytes\\
DNA Sensor Actor Interface  & 41 kBytes  & 22 kBytes\\
AHS                         & 146 kBytes & 379 kBytes\\
                            &            & \\
Processor Overall           & 270 kBytes & 1662 kBytes\\
\noalign{\smallskip}\hline
\end{tabular}
\end{table}

\section{Evaluation}
\label{sec:evaluation}

\begin{figure}[ht!]
     \centering
	   \includegraphics[width=0.8\linewidth]{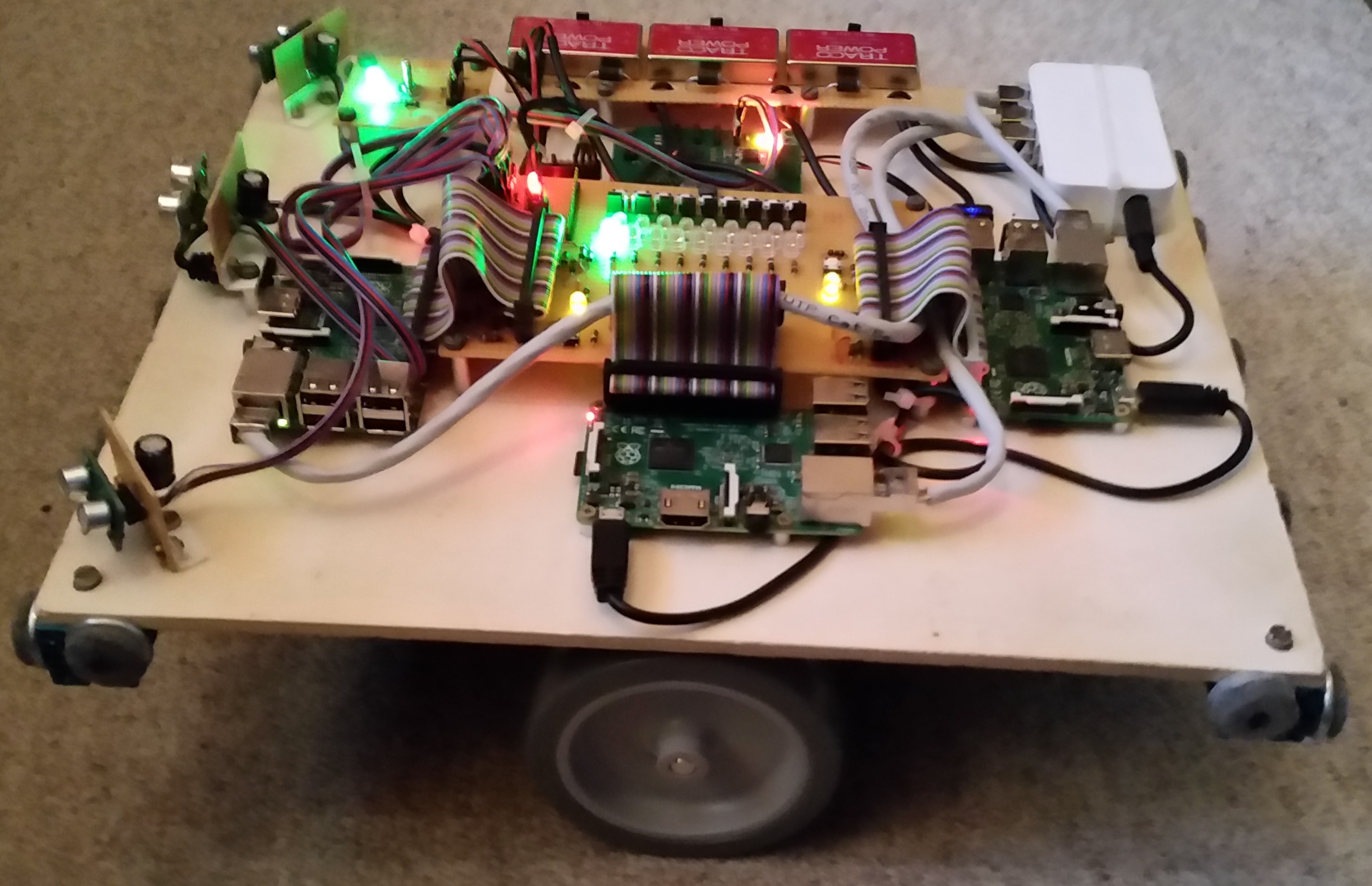}
	   \caption{The demonstrator vehicle}
	   \label{fig:DemVehicle}
\end{figure}

\begin{figure}[ht!]
     \centering
	   \includegraphics[width=0.8\linewidth]{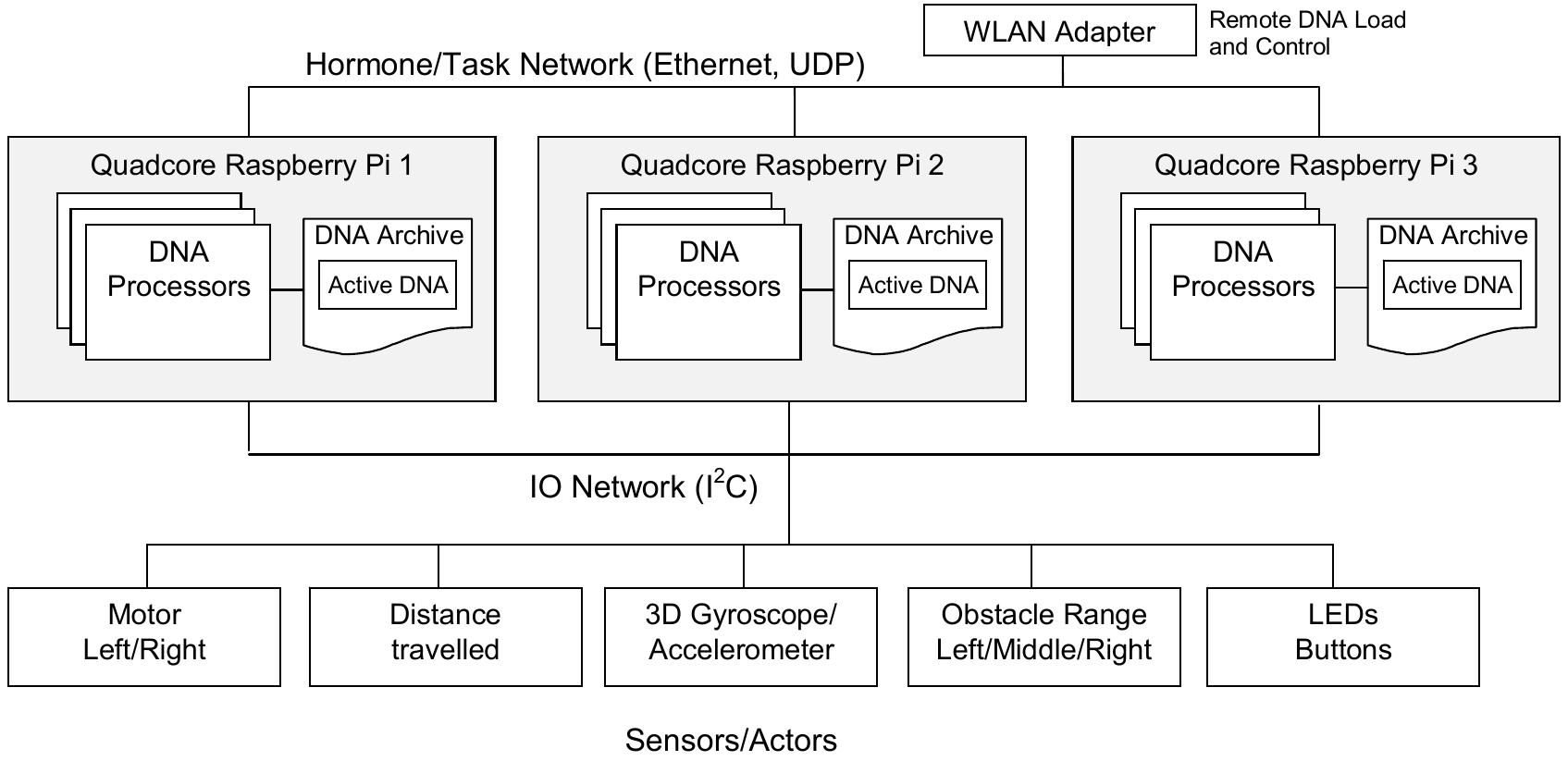}
	   \caption{Architecture of the demonstrator vehicle}
	   \label{fig:DemArch}
\end{figure}

A first evaluation result has already been presented in the previous section. The memory footprint of the Artificial DNA implementation is rather small as shown in table \ref{table:components}. To conduct further evaluations, we have chosen a flexible robot vehicle platform (FOXI)\footnote{\textbf{F}lexible \textbf{O}rganix e\textbf{X}perimental veh\textbf{I}cle} as a demonstrator. This platform can be used either as a self-balancing two wheel vehicle (as an inverse pendulum like e.g. a Segway personal transporter) or a stable four wheel vehicle by two additional foldaway supporting wheels. It uses a differential drive and is equipped with various sensors and actors. This allows a wide range of real-time applications. Figure \ref{fig:DemVehicle} shows a picture of the vehicle (while running a self-balancing application DNA) and figure \ref{fig:DemArch} sketches the vehicle architecture. It holds three quadcore Raspberry Pi processors running Raspian Linux. Three cores of each Pi are used as DNA processors resulting in overall 9 DNA processors on the demonstrator platform\footnote{Since Raspian Linux is no real-time OS, the fourth core of each Pi is spared for operating system usage.}. The Pis are interconnected by Ethernet. Additionally, a WLAN module is connected. This allows to load DNA files from outside to the DNA archive of each Pi and to remotely control which DNA is cloned (loaded) from the DNA archive to all DNA processors. It is guaranteed that all DNA processors (the cells) use the same DNA at a time\footnote{However, the DNA on all processors can be changed at run-time simultaneously.}. Furthermode, additional external DNA processors and external sensors and actors e.g. on a Windows PC can be attached via WLAN to extend the demonstrator\footnote{As a restriction, external DNA processors have no access to the internal sensors and actors of the vehicle, so basic elements attached to these sensors and actors will not be allocated to external DNA processors. Furthermore, real-time capabilities of external DNA processors, sensors and actors are limited due to the WLAN connection.}. All internal sensors and actors are attached to and shared by the Raspberry Pis via an I$^2$C bus\footnote{since the native I$^2$C bus interface of the Raspberry Pi is not multi-master capable, we added some additional hardware support to realize multi-master access.}. Available sensors are three supersonic rangefinders (to detect obstacles left, right or in front of the vehicle in autonomous driving applications), a three axis gyroscope and accelerometer (used e.g. for self-balancing applications), an odometer for the left and the right drive wheel (to measure the distance traveled) and several push buttons as digital inputs. Actors are the left and right motor of the differential drive and several LEDs which can be used as digital outputs or dimmed by PWM. The power supply of each Pi can be shutdown remotely or by a push button to inject a heavy component failure (turning off the power of one Raspberry Pi means the simultaneous hard failure of three DNA processors). Single cores can be shutdown individually as well\footnote{Failures of the communication system are not in the focus of this demonstrator. This will be covered by future research work.}.

\begin{figure}[ht!]
     \centering
	   \includegraphics[width=0.8\linewidth]{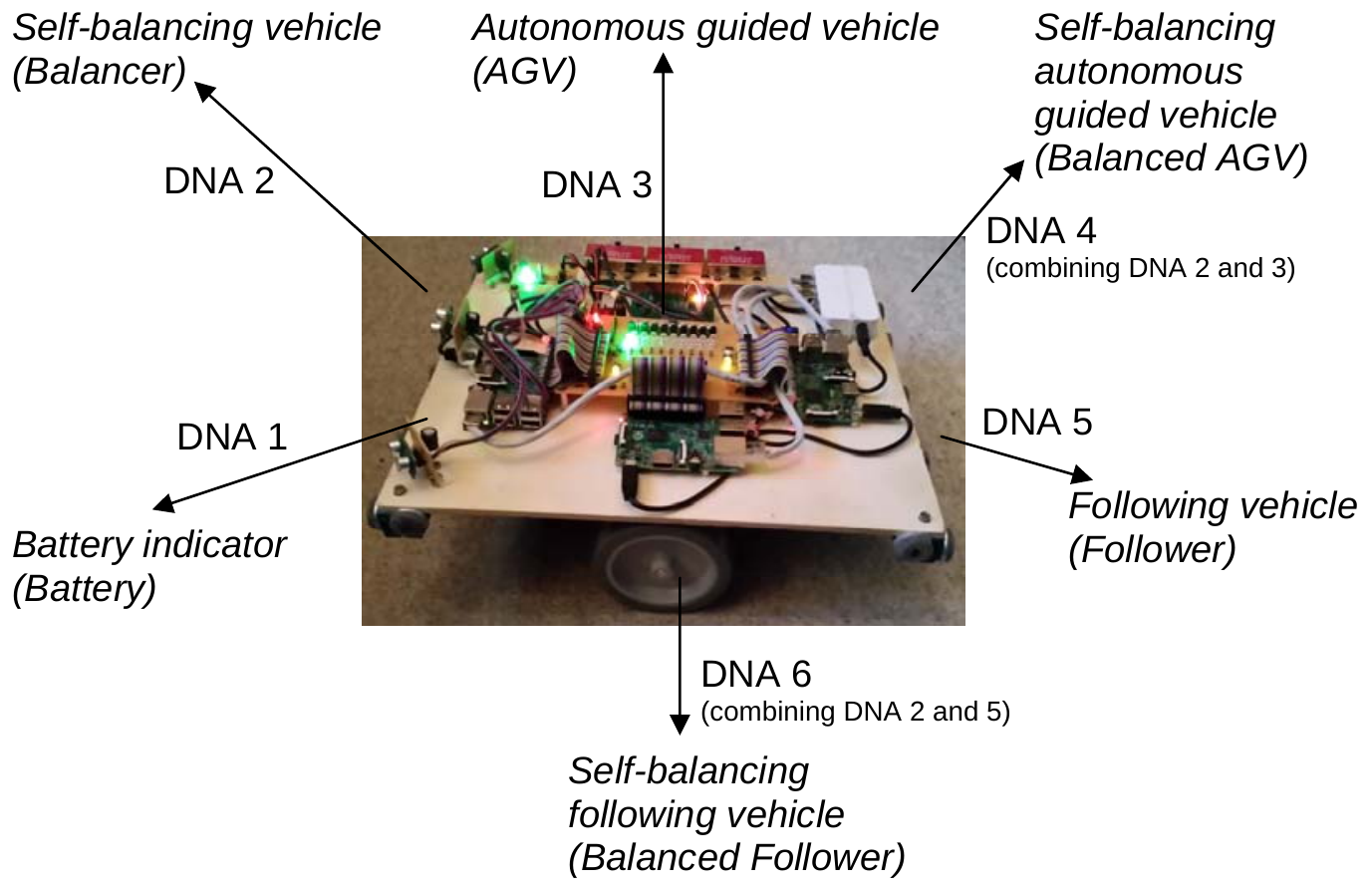}
	   \caption{Shaping the vehicle by different DNAs}
	   \label{fig:VehicleDNAs}
\end{figure}

Loading a specific DNA to the demonstrator vehicle platform now determines what the vehicle will become. For evaluation purpose we have created several different DNAs as shown in Figure \ref{fig:VehicleDNAs}.

A demonstration video showing experiments with these DNAs can be found at \cite{brinksvideo}.

\subsection{A First Example}

\begin{figure*}[ht!]
     \centering
	   \includegraphics[width=0.8\linewidth]{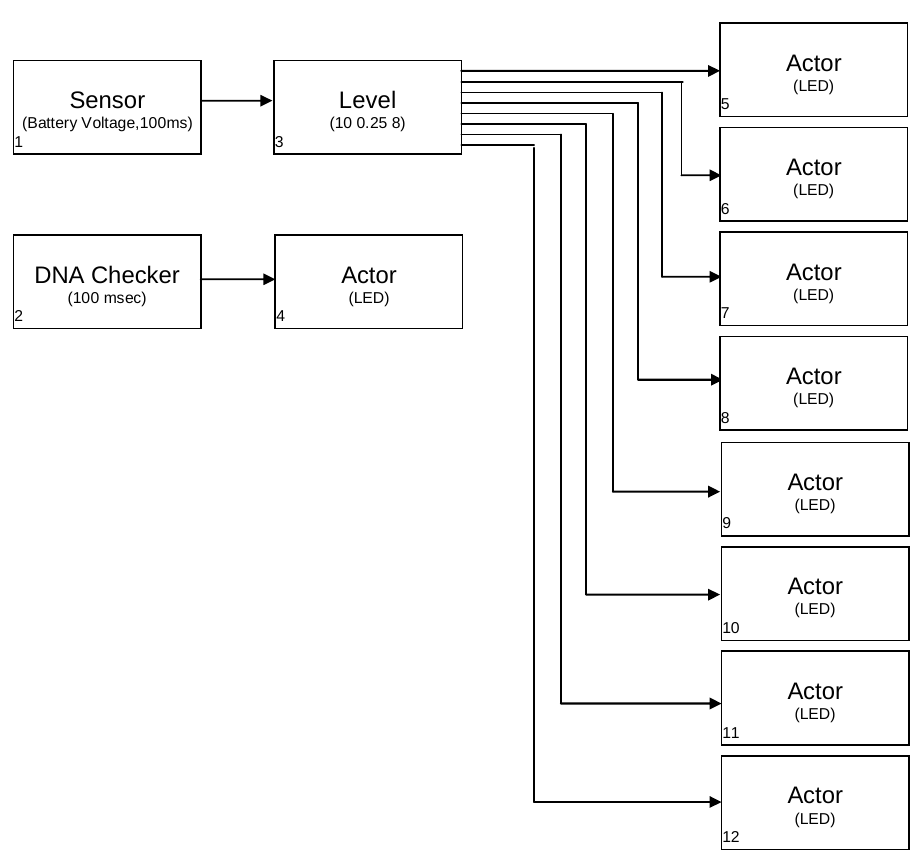}
	   \caption{Block diagram of the Battery DNA}
	   \label{fig:batteryDNA}
\end{figure*}

As a first very simple example, Figure \ref{fig:batteryDNA} shows the block diagram of a DNA for a battery indicator. It displays the battery voltage by a bar of LEDs. For better identification of the basic elements in this figure they are numbered in the left lower corner. The DNA consists of a battery voltage sensor (basic element 1) which delivers its output each 100 milliseconds to a level discriminator (basic element 3). This discriminator triggers actor LEDs (basic elements 5 - 12) depending on the input voltage level. The discriminator levels are defined by the parameter set $10$ (high voltage) $0.25$ (voltage step) and $8$ (low voltage). Finally, another LED actor (basic element 4) shows when the appliction has been completely built or rebuilt (in case of a failure) by its DNA using the DNA Checker basic element (2) with a period of also 100 milliseconds. The DNA consist of 12 basic elements and uses only 4 different basic elements. The size of the DNA stored in the compact form proposed in Section \ref{ssec:artDNA} and \cite{brinksCandC} is 162 Bytes.

\begin{figure*}[ht!]
     \centering
	   \includegraphics[width=0.8\linewidth]{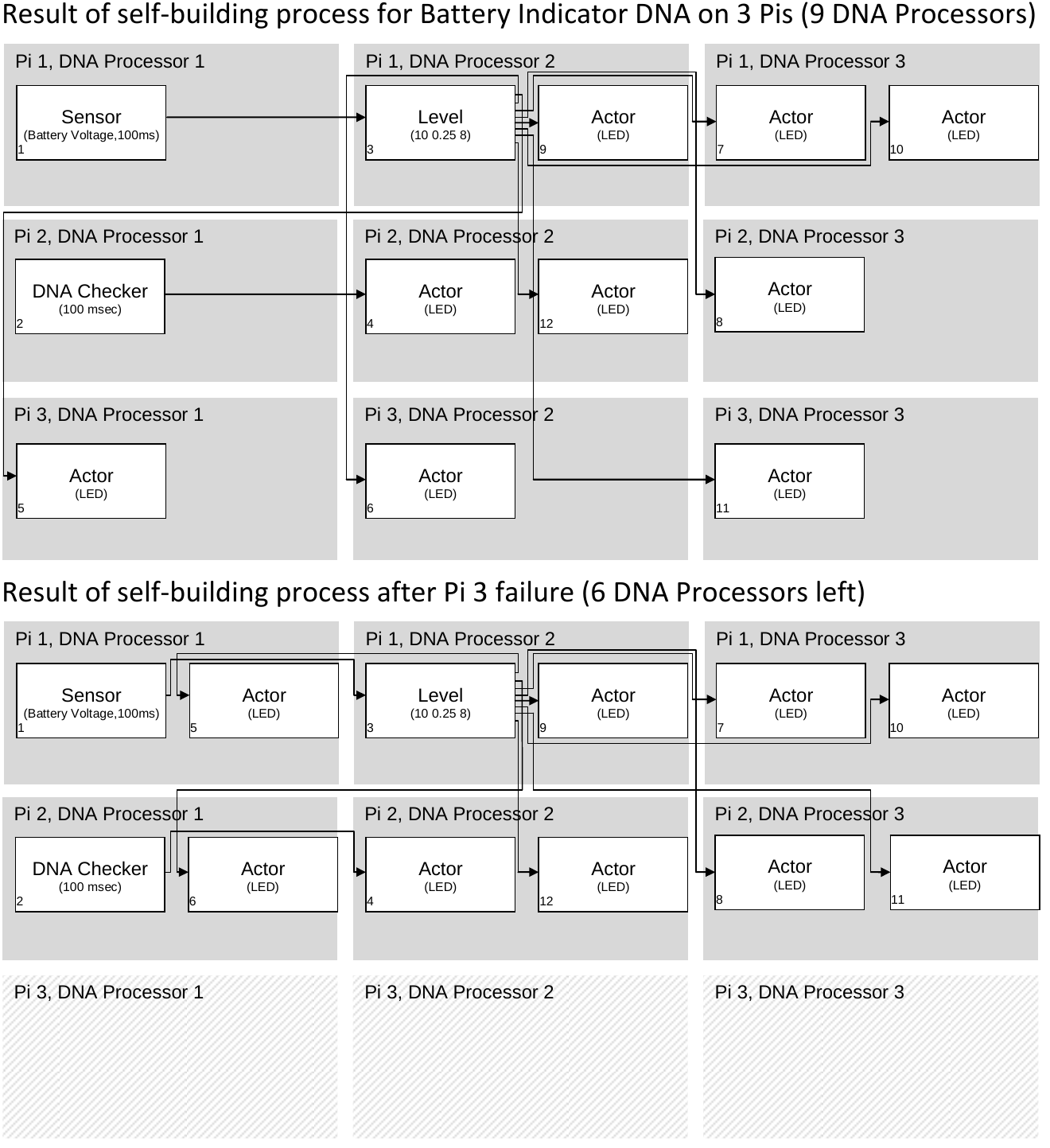}
     \caption{Allocation and connection of Battery Indicator basic elements before and after killing of Pi 3}
	   \label{fig:batteryProcessors}
\end{figure*}

\begin{figure*}[ht!]
     \centering
	   \includegraphics[width=0.8\linewidth]{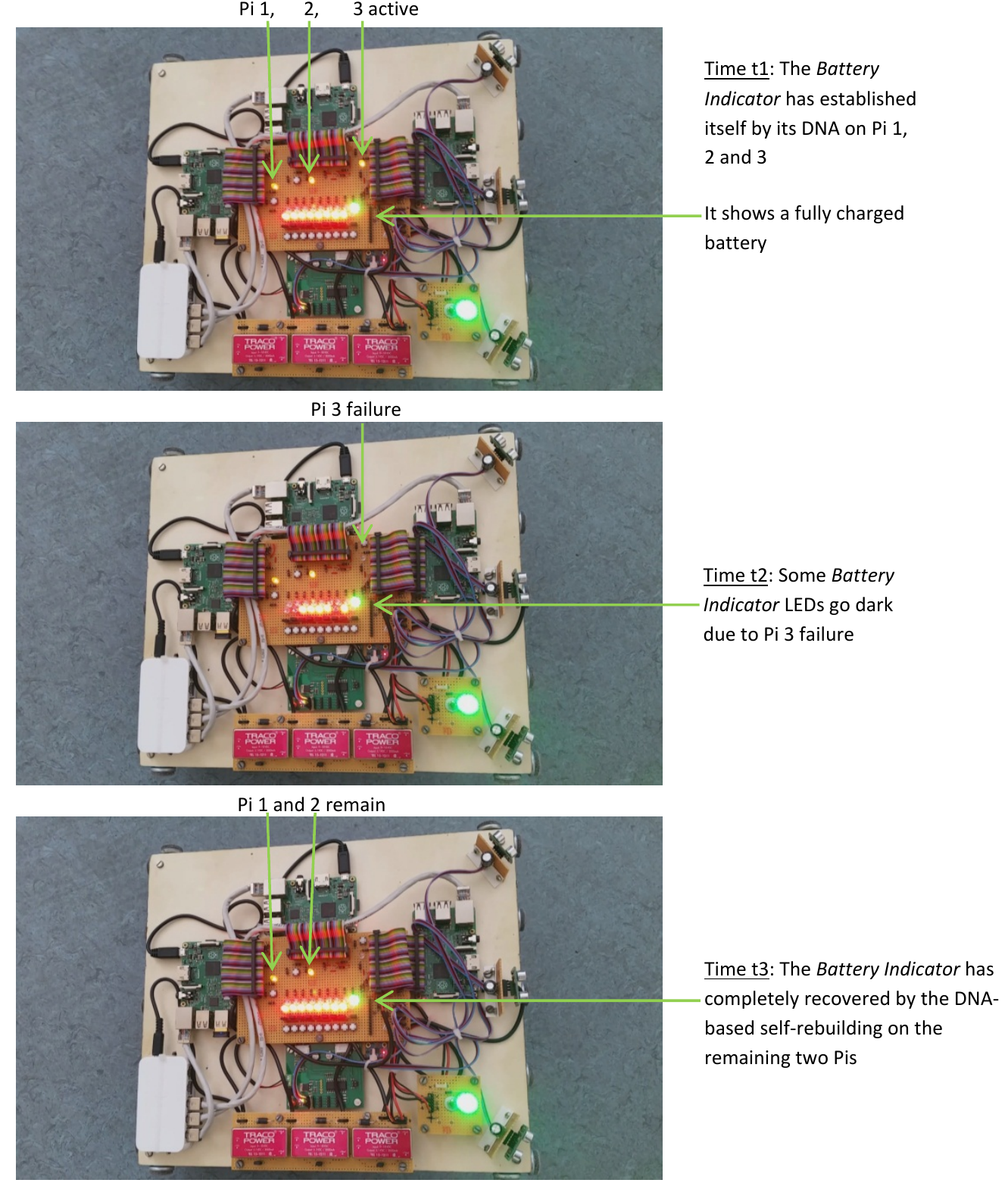}
	   \caption{DNA-based self-healing of the Battery Indicator}
	   \label{fig:batteryScreenshots}
\end{figure*}

This simple DNA can be used as a good first example to demonstrate the self-healing properties of the concept in a qualitive way. When loading the DNA and starting the system, the battery indicator builds itself based on its DNA. The upper part of Figure \ref{fig:batteryProcessors} shows how the battery indicator initially allocates and connects itself to the 9 DNA processors of the 3 Raspberry Pis. The upper picture of Figure \ref{fig:batteryScreenshots} shows a snapshot of the vehicle once the battery indicator has established itself at time $t1$. The bar of red LEDs shows the battery is complelety charged. The three yellow LEDs above the bar indicate all 3 Rapberry Pis are powered up and active. At time $t2$ we suddenly shut down the power supply of Raspberry Pi 3. This results in the simultaneous failure of its 3 DNA processors, which hold 3 LED actors (basic elements 5, 6 and 11, see upper part of Figure \ref{fig:batteryProcessors}). So for a very short moment those 3 red LEDs go dark as can be seen in the middle snapshot picture of Figure \ref{fig:batteryScreenshots}. Furthermore, the rightmost yellow LED is now dark since Raspberry Pi 3 is down. However, based on its DNA the battery indicator rebuilds itself autonomously. At time $t3$ the system has completely recovered as can be seen in the lower snapshot picture of Figure \ref{fig:batteryScreenshots}. This happens very quickly, as the green LED (basic element 4, located right next to the red LED bar) which checks and indicates the completeness of the DNA every 100 milliseconds, doesn't even go dark. The lower part of Figure \ref{fig:batteryProcessors} shows the result of reallocation and reconnection to the remaining 6 DNA processors. Actually, basic elements 5, 6 and 11 have autonomously moved to new DNA processors, as also indicated in table \ref{table:tasksBattery}. As mentioned in section \ref{sec:build}, the time complexity of this rebuilding process is $\mathcal O(n)$. With $n$ equal to 3 basic elements to relocate, this is achieved in less than 150 msec, so the flickering of the LED is too short to be visible.

\begin{table}[ht!]
\centering
\caption{Movement of Battery Indicator basic elements due to killing of Pi 3}
\label{table:tasksBattery}
\begin{tabular}{c | c | c | c }
\hline\noalign{\smallskip}
Pi & DNA       & Basic    & Basic\\
   & Processor & Elements & Elements\\
   &           & before killing Pi 3 & after killing Pi 3\\
\noalign{\smallskip}\hline\noalign{\smallskip}
1 & 1 & 1 & 1, \textit{\textbf{5}}\\
1 & 2 & 3, 9 & 3, 9\\
1 & 3 & 7, 10& 7, 10\\
2 & 1 & 2 & 2, \textit{\textbf{6}}\\
2 & 2 & 4, 12 & 4, 12\\
2 & 3 & 8, & 8, \textit{\textbf{11}}\\
3 & 1 & 5& \\
3 & 2 & 6& \\
3 & 3 & 11& \\
\noalign{\smallskip}\hline
\end{tabular}
\end{table}

\subsection{More DNAs}
\label{sec:moreDNAs}

\begin{figure*}[ht!]
     \centering
	   \includegraphics[width=0.8\linewidth]{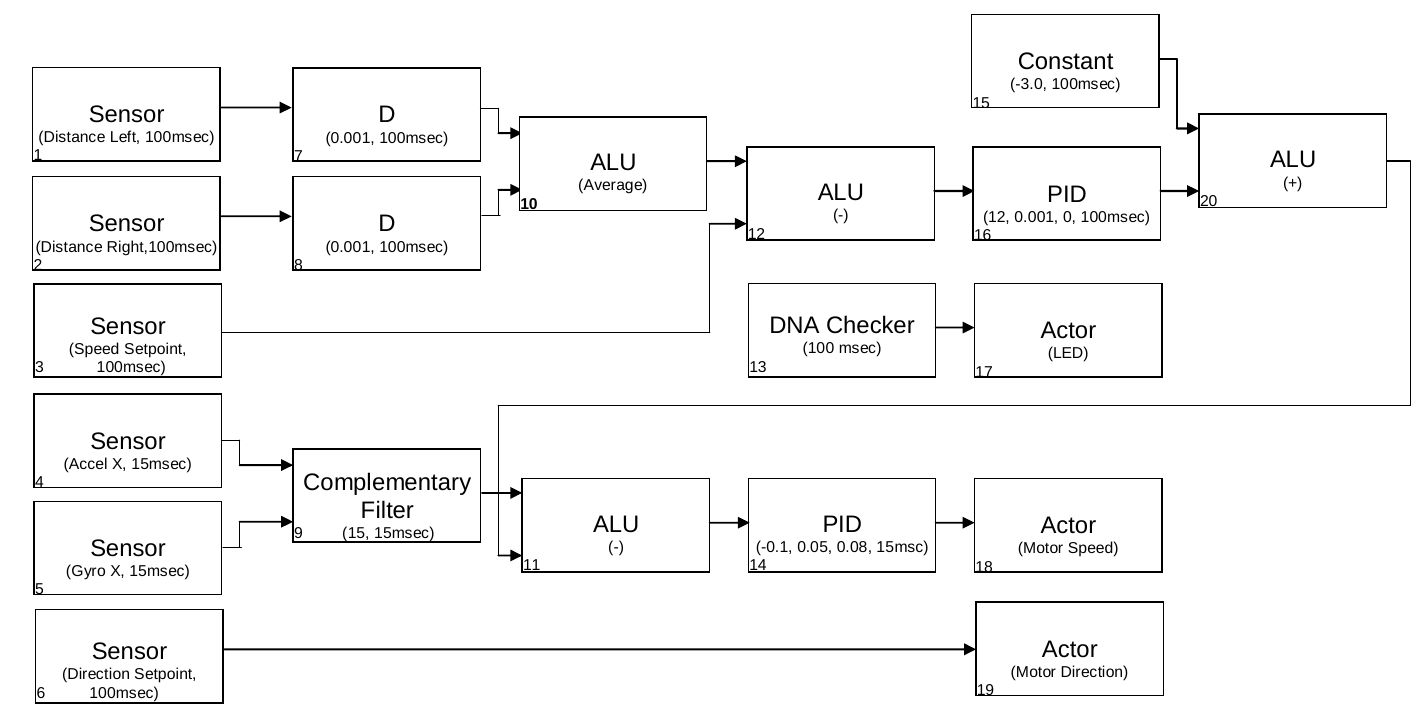}
	   \caption{Block diagram of the Balancer DNA}
	   \label{fig:balancerDNA}
\end{figure*}

In the following, a more detailed quantitative evaluation is conducted based on more complex DNAs.
Figure \ref{fig:balancerDNA} shows a DNA that makes the vehicle self-balancing like e.g. a Segway. Like before, the basic elements in this figure are numbered in the left lower corner. The self-balancing DNA basically consists of a cascaded closed control loop. The outer loop (basic elements 1, 2, 3, 7, 8, 10, 12, 16) controls the speed of the vehicle by a PID controler. The current speed is determined by differentiating (7, 8) and averaging (10) the odometer data of the left (1) and right (2) wheel of the differential drive. The desired speed setpoint is read by an external sensor (3) via WLAN. A PID controller (16) sets the vehicle angle by the speed deviation (12) with a period of 100 milliseconds. If the vehicle is too slow it will tilt forward to accelerate. If the vehicle is too fast it will tilt backward to slow down. With a slight correction regarding the mass center (15, 20) this is the setpoint for the inner loop which controls the vehicle angle (basic elements 4, 5, 9, 11, 14, 18). The current angle is determined by the fusion of accelerometer (4) and gyroscope (5) data using a complementary filter (9). This is necessary because pure accelerometer data is noisy and gyroscope data has a permanent drift. A PID controller (14) accelerates or decelerates the differential drive (18) using the angle deviation (11) with a period of 15 milliseconds to achive the desired angle. The desired direction of the vehicle is read by another external sensor (6) via WLAN and is directly connected to the direction actor (19) of the differential drive. Finally, a LED (17) shows when the appliction has been completely built or rebuilt (in case of a failure) by its DNA using the DNA Checker basic element (13). This DNA consists of 20 basic elements and uses 8 different basic elements. The size of this DNA stored in the compact form is 188 Bytes. It shows the artifical DNA concept allows a very compact representation of embedded applications.

\begin{figure*}[ht!]
     \centering
	   \includegraphics[width=0.8\linewidth]{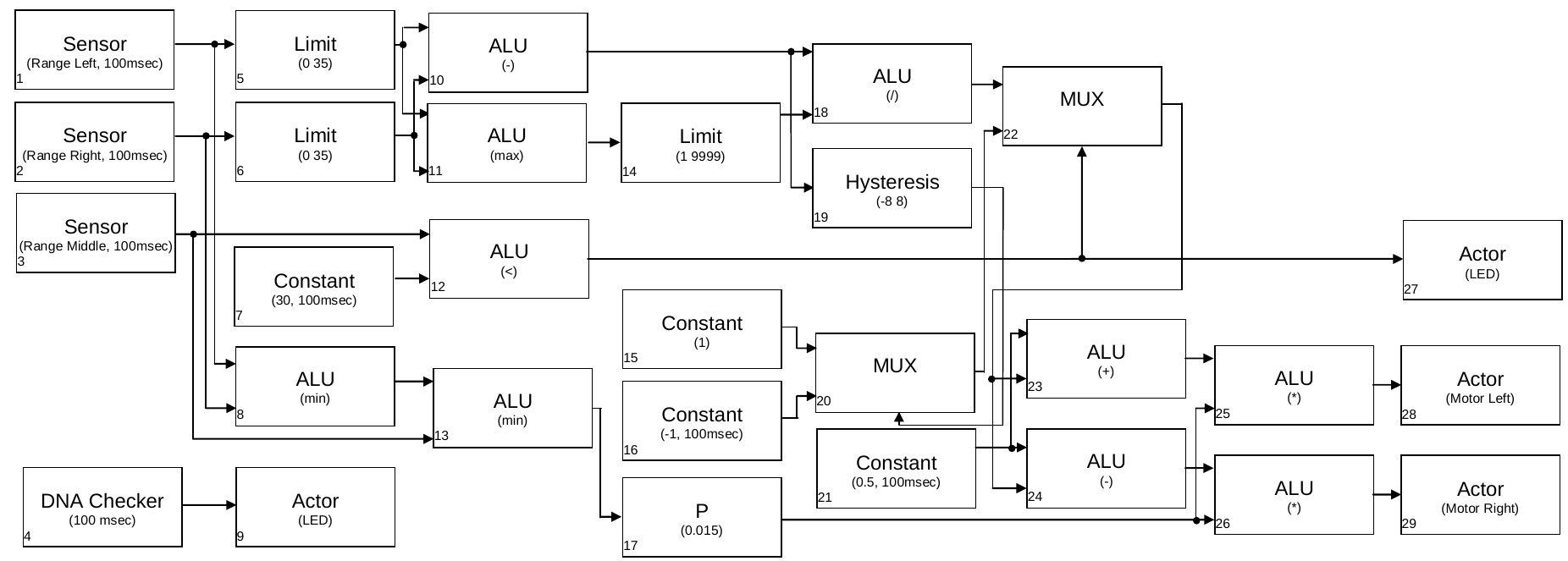}
	   \caption{Block diagram of the AGV DNA}
	   \label{fig:agvDNA}
\end{figure*}

To further evaluate this, we have created more DNAs: An \textit{Autonomous Guided Vehicle (AGV) DNA} shown in Figure \ref{fig:agvDNA} autonomously drives the vehicle in a maze using the supersonic range finder sensors. This DNA uses the supporting wheels so no self-balancing is necessary. Based on the left and right range finders (basic elements 1, 2) a driving direction is calculated (5, 6, 10, 11, 14, 18). In case of an obstacle very close to the mid range finder (basic element 3), an evasive turn action is provided (7, 12, 15, 16, 19, 20, 22). The vehicle speed is calculated by the lowest value of all three range finders (8, 13, 17). This DNA applies direct control to the left and right motor of the differential vehicle drive (23, 24, 25, 26, 28, 29).

\begin{figure*}[ht!]
     \centering
	   \includegraphics[width=0.8\linewidth]{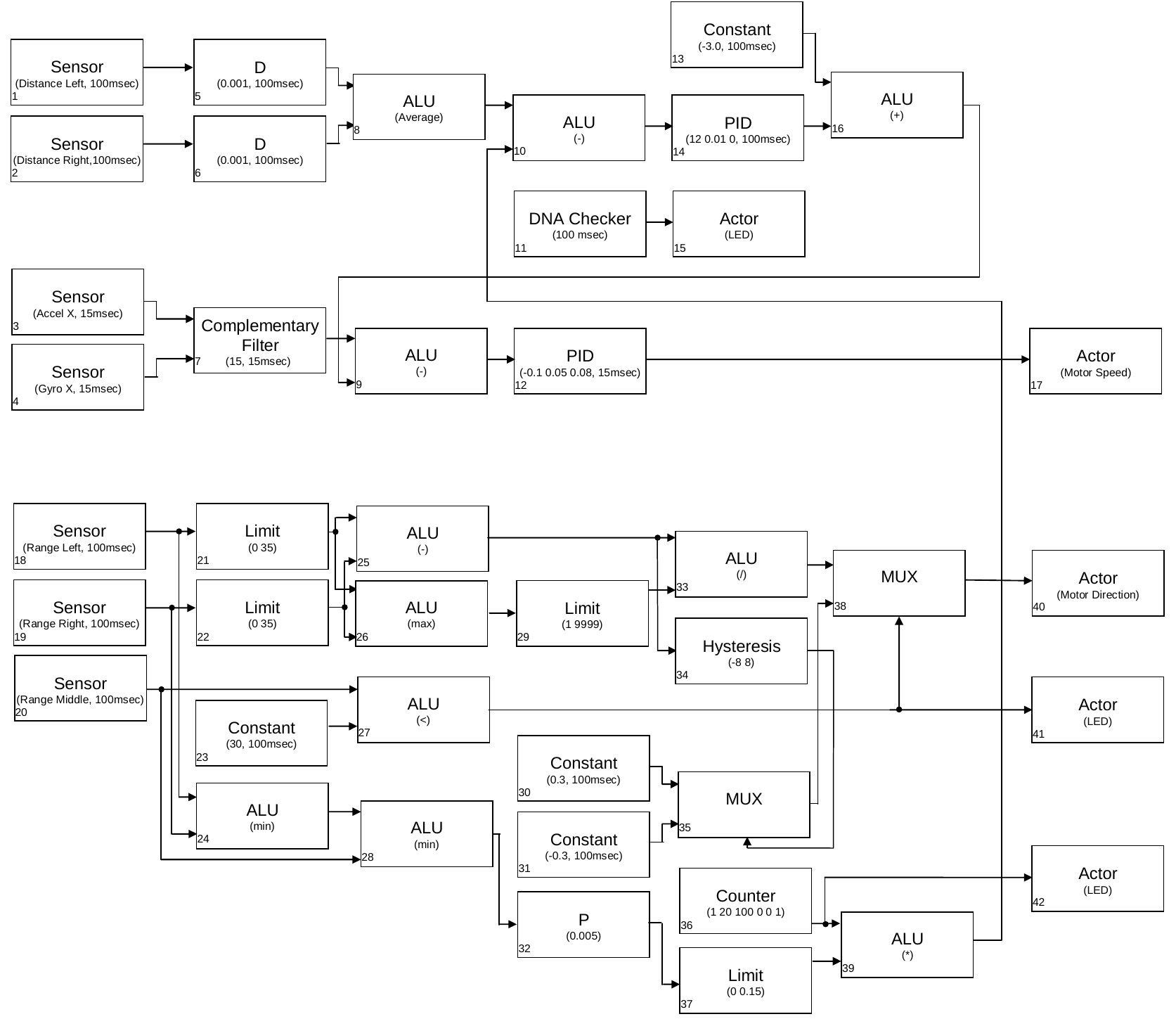}
	   \caption{Block diagram of the Balanced AGV DNA}
	   \label{fig:balanceagvDNA}
\end{figure*}

A \textit{Balanced AGV DNA} sketched in Figure \ref{fig:balanceagvDNA} combines the self-balancing (basic elements 1 - 17) and the AGV DNA (basic elements 18 - 42) to create a self-balancing autonomous vehicle. 

\begin{figure*}[ht!]
     \centering
	   \includegraphics[width=0.8\linewidth]{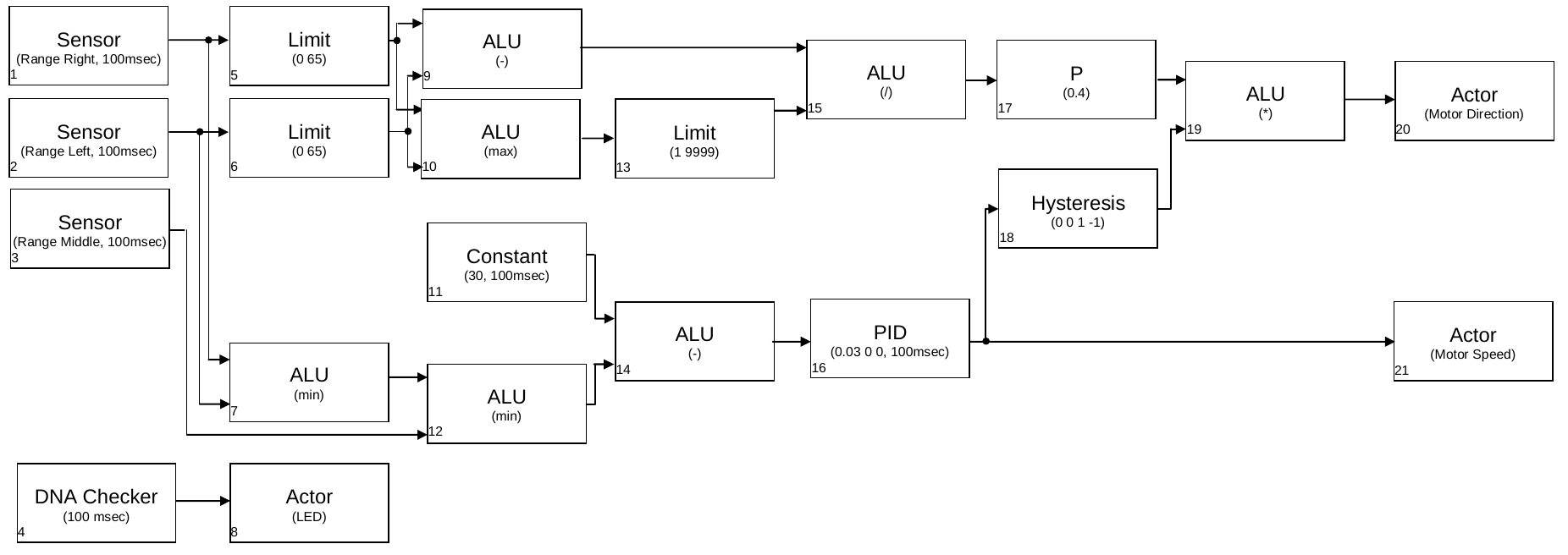}
	   \caption{Block diagram of the Follower DNA}
	   \label{fig:followDNA}
\end{figure*}

A \textit{Follower DNA} displayed in Figure \ref{fig:followDNA} lets the vehicle follow an object using the rangefinders. The direction to the obstacle is calculated by the left and right range finders (1, 2, 5, 6, 9, 10, 13, 15, 17, 19, 20). The speed is calculated by a closed PID control loop (14, 16, 21) to keep it at a desired distance of 30 cm (11) from the obstacle. The distance is derived by the miminum of all thre range finders (1, 2, 3, 7, 12).

\begin{figure*}[ht!]
     \centering
	   \includegraphics[width=0.8\linewidth]{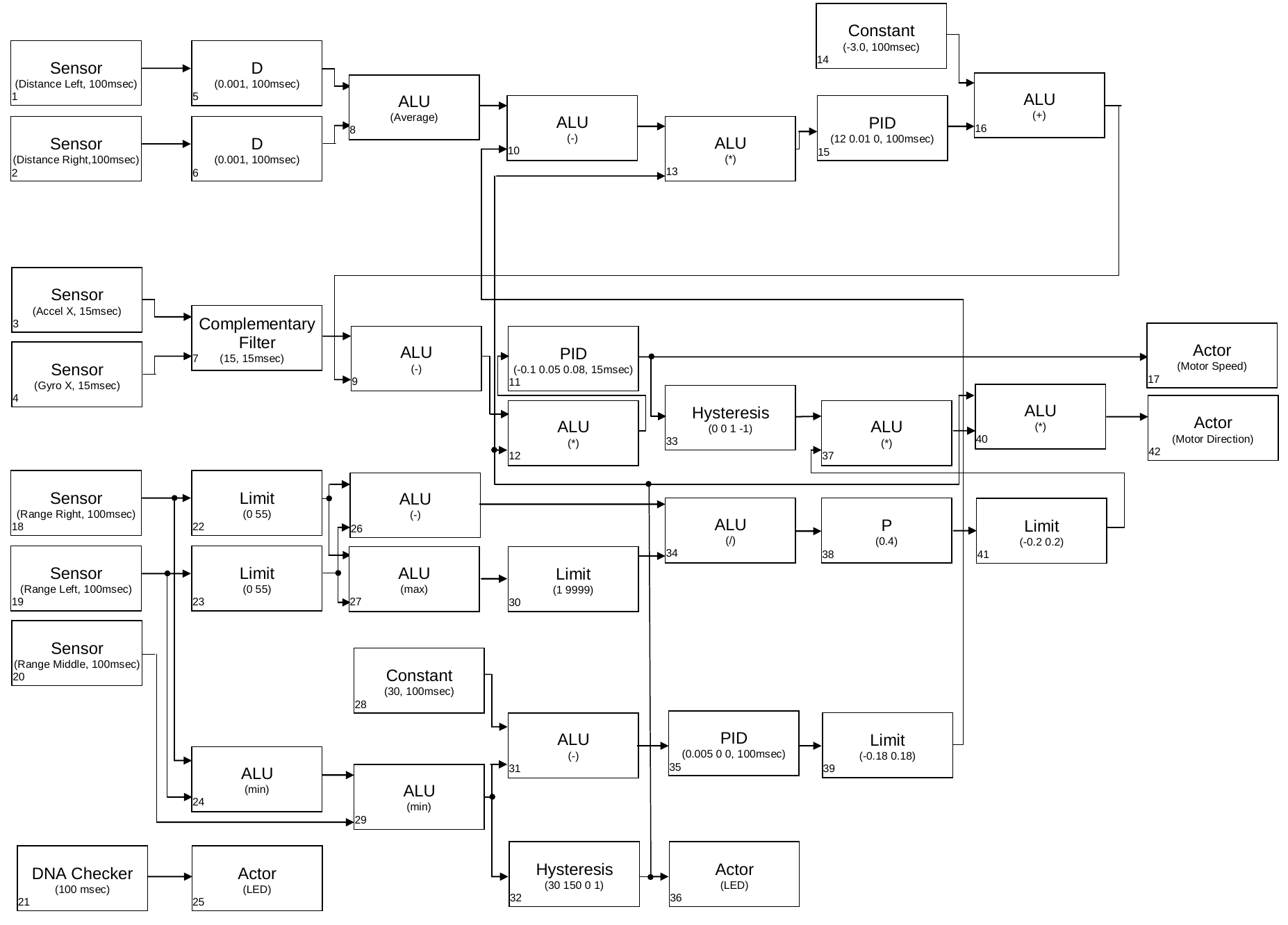}
	   \caption{Block diagram of the Balanced Follower DNA}
	   \label{fig:balancefollowerDNA}
\end{figure*}

Finally, a \textit{Balanced Follower DNA} shown in Figure \ref{fig:balancefollowerDNA} combines the self-balancing (basic elements 1 - 17) and the Follower DNA (basic elements 18 - 42) to create a self-balancing follower. 

Table \ref{table:sizes} gives the sizes of these DNAs. They are all very small and only consist of few different basic elements. Furthermore, the load produced by the communication of the basic elements for each application is given in this table. This load is also considerably small in the range of some kBytes per second.       

\begin{table}[ht!]
\centering
\caption{Sample DNAs}
\label{table:sizes}
\begin{tabular}{l | c | c | c | c}
\hline\noalign{\smallskip}
DNA                         & Basic  & Different & DNA  & Communi-\\
                            & Elements & Basic & Size & cation Load\\
                            &        & Elements & (Bytes)     & (Bytes/sec)\\
\noalign{\smallskip}\hline\noalign{\smallskip}
Battery            & 12 & 4  & 162 & 2100\\
Balancer           & 20 & 8  & 188 & 9513\\
AGV                & 29 & 10 & 338 & 7140\\
Balanced AGV       & 42 & 13 & 536 & 15183\\
Follower           & 21 & 10 & 270 & 5040\\
Balanced Follower  & 42 & 11 & 536 & 22953\\
\noalign{\smallskip}\hline
\end{tabular}
\end{table}

\subsection{Real-time Behavior}

The basic elements are allocated and connected to the DNA processors by the DNA and the AHS in a self-organizing way. Table \ref{table:tasks} shows this exemplarily for the self-balancing DNA. The first row shows the initial allocation to the available 9 DNA processors. After a while the power supply of Raspberry Pi 2 was shut down turning off tree DNA processors simultaneously. The DNA and AHS now reallocate and reconnect the basic elements so the remaining 6 DNA processors can still perform the application. The new allocation is shown in row 2 of the table. The reallocated basic elements are marked in italic. 

\begin{table}[ht!]
\centering
\caption{Balancer: allocation of basic elements before and after killing of Pi 2}
\label{table:tasks}
\begin{tabular}{c | c | c | c }
\hline\noalign{\smallskip}
Pi & DNA       & Basic    & Basic\\
   & Processor & Elements & Elements\\
   &           & before killing Pi 2 & after killing Pi 2\\
\noalign{\smallskip}\hline\noalign{\smallskip}
1 & 1 & 13, 17 & \textit{\textbf{6}}, 13, 17, \textit{\textbf{19}}\\
1 & 2 & 4, 5, 9, 11 & 4, 5, 9, 11\\
1 & 3 & 15, 20& \textit{\textbf{12}}, 15, \textit{\textbf{16}}, 20\\
2 & 1 & 6, 19&\\
2 & 2 & 10, 12, 16&\\
2 & 3 & 3&\\
3 & 1 & 14, 18& \textit{\textbf{3}}, 14, 18\\
3 & 2 & 1, 7& 1, 7, \textit{\textbf{10}}\\
3 & 3 & 2, 8& 2, 8\\
\noalign{\smallskip}\hline
\end{tabular}
\end{table}

\begin{figure}[ht!]
     \centering
	   \includegraphics[width=0.8\linewidth]{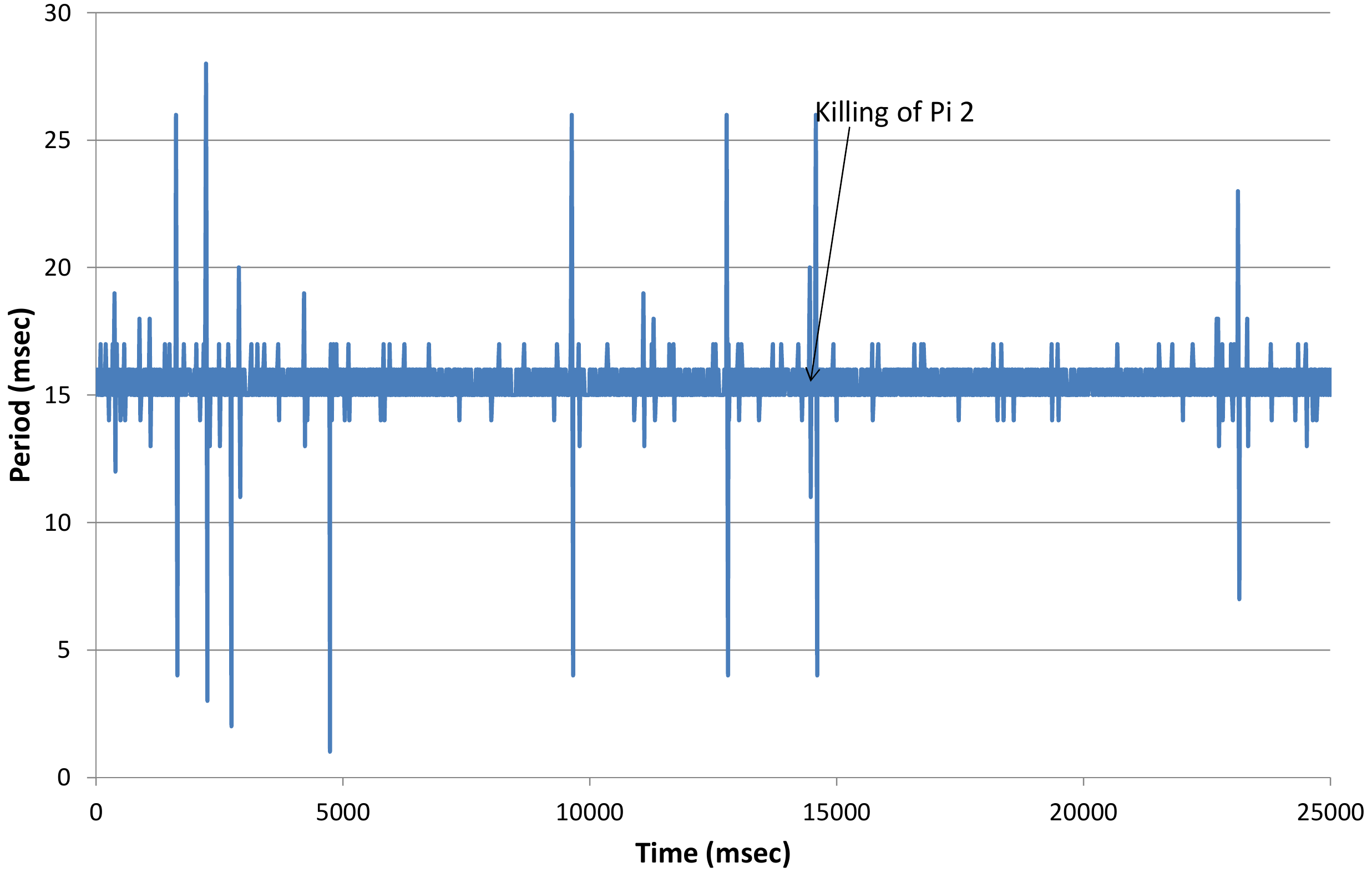}
	   \caption{Real-time properties: period of the motor control signal}
	   \label{fig:diag2}
\end{figure}

\begin{figure}[ht!]
     \centering
	   \includegraphics[width=0.8\linewidth]{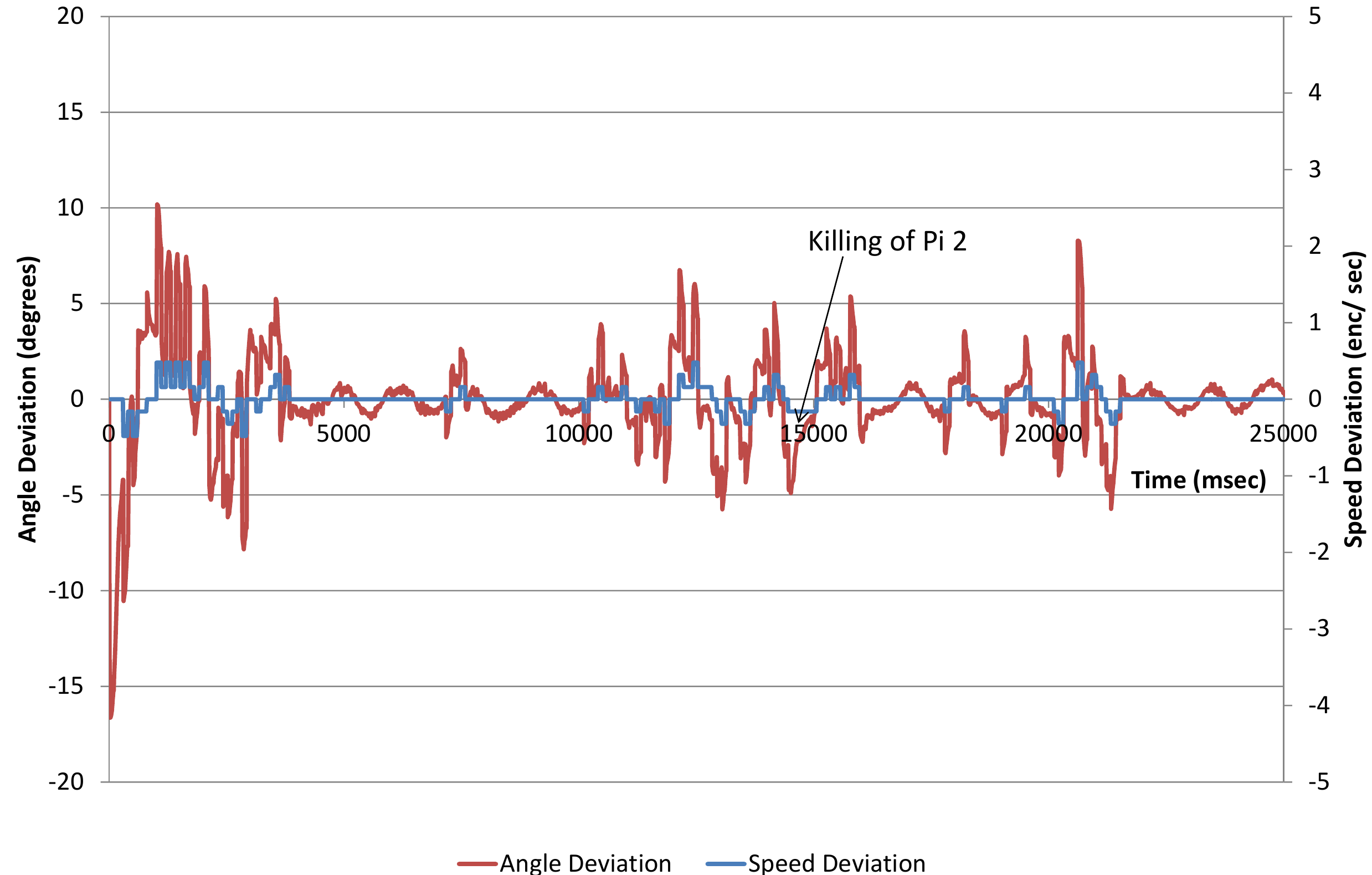}
	   \caption{Angle and speed deviation of the vehicle controlled by the balancer DNA}
	   \label{fig:diag1}
\end{figure}

\begin{figure}[ht!]
     \centering
	   \includegraphics[width=0.8\linewidth]{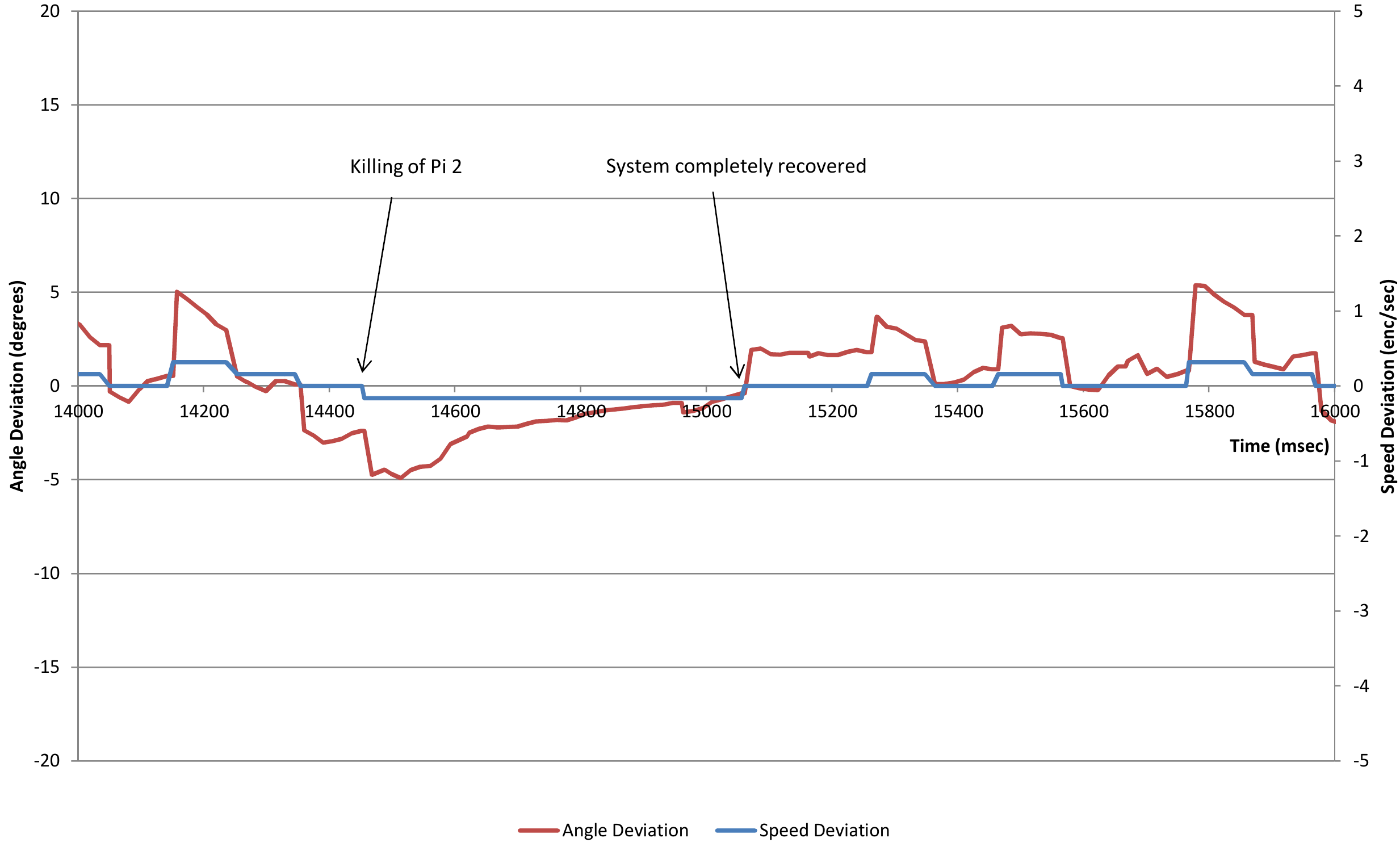}
	   \caption{Extract of Figure \ref{fig:diag1}}
	   \label{fig:diag1a}
\end{figure}

\begin{figure}[ht!]
     \centering
	   \includegraphics[width=0.8\linewidth]{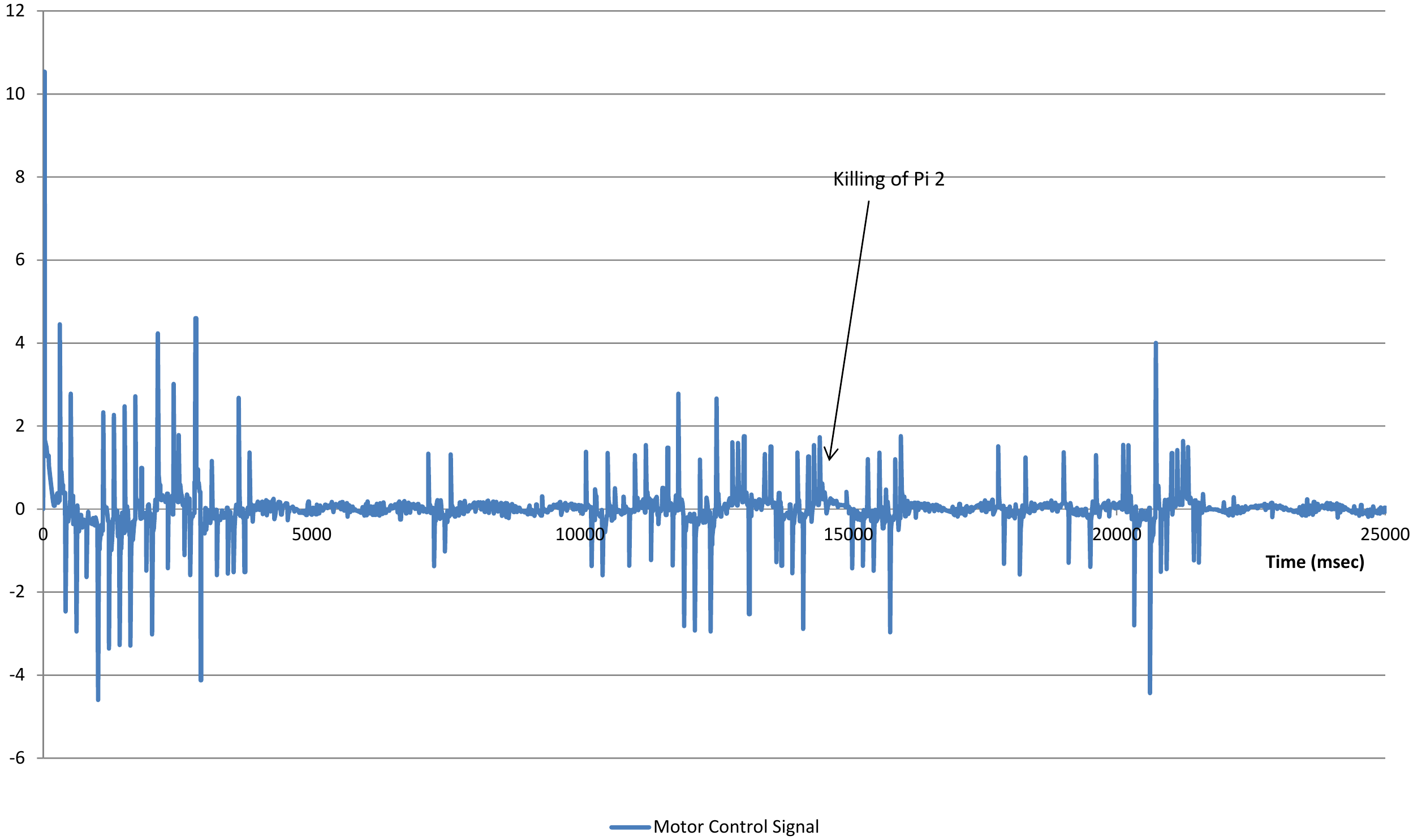}
	   \caption{Motor control signal}
	   \label{fig:diag3}
\end{figure}

We use this DNA scenario to evaluate the real-time behavior of the Artificial DNA system on the demonstrator platform. The results of the other DNAs are very similar. Figure \ref{fig:diag2} shows the measured period of the motor control signal (output of basic element 14 in the balance DNA). As mentioned above the period of the inner control loop is set to 15 msec by the DNA parameters. Overall this period is reached quite well, however Raspian Linux is no real-time operating system. So some occasional spikes in the range of plus/minus 12 msec to the intended period can be observed. It is very interesting that shutting down Pi 2 at timestamp 14800 only produces a spike of plus/minus 10 msec, which is in the range of the other spikes. So the self-rebuilding of the system by the DNA works almost seamlessly and fast. The vehicle does not loose its balance. This can be seen in figure \ref{fig:diag1}. Here, the deviation of the desired speed and angle of the vehicle are shown. After the initial self-building of the system which takes about 700 msec, the target angle and speed are well reached. The small angle deviations of about plus/minus 5 degrees result from the remaining noise of the accelerometer, friction, mechanical play of the drive and the occasional spikes in the control period as shown in the previous diagram. Interestingly the shutdown of Pi 2 does not cause a major disturbance in angle and speed control. This is shown in more detail in figure \ref{fig:diag1a}. The deviation is in the range of the other deviations. The rebuilding process is fast enough to keep well the balance and the speed. Additionally, Figure \ref{fig:diag3} shows the motor control signal itself. Also here it can be seen that the shutdown of Pi 2 does not cause any extraordinary change in this signal.

Overall, the evaluation shows the concept of the artificial DNA well enables the self-building, self-adapting and self-restoring of real-time applications at run-time with respect to the currently available hardware platform. Memory sizes and communication overhead are considerably low.

\section{Future Work}
\label{sec:futurework}
The prototypic implementation presented and evaluted in the previous two sections is adaptive in the way that a building plan given by an artificial DNA adapts itself as good as possible to the current hardware platform. The builing plan itself stays unchanged. However, the concept of the artificial DNA offers more possibilities for adaptation. By modifying the artificial DNA at run-time the building plan can also adapt or optimze itself to changing requirements. There are two possibilities for such an adaptation: In case of \textit{reconfiguraton} the modification of the artificial DNA is initiated by a direct user interaction or by a trigger condition defined by the user. Then, predefined parts of the artificial DNA will be switched on or switched off or be replaced by also predefined new parts. In case of \textit{evolution} the system itself autonomously explores modifications of the artificial DNA. If a modification improves the system behavior it will be kept, otherwise it will be discarded. The type and scale of such modifications are determined by the system, but can be guided by constraints, rules and goals given by the user to control the evolutionary process.

The self-healing capabilities of the prototypic implementation are mentioned to compensate core failures. The artificial DNA also allows more \textit{fine grained fault-tolerance mechanisms which can be synthesized at run-time}. Depending on the importance of artificial DNA sequences for an application and the availability of system resources, critical parts can be automatically replicated at run-time. The parts replicated and the number of replicas (double modular redundancy, triple modular redundancy, ...) can be autonomously adapted to changing environmental conditions. Since the complete building plan of the application including all dependencies is present in each processor core by the artificial DNA, fault detection and correction measures can be flexibly applied.  

\subsection{Reconfiguration}

A reconfiguration of the artificial DNA initiated by a direct user interaction allows to modify and update an application without stopping the system. This is important for many embedded real-time applications. Basic operations for such a reconfiguration are the \textit{change of parameters} of basic elements, the \textit{removement of artificial DNA sequences},
the \textit{insertion of artificial DNA sequences} and the \textit{replacement of artificial DNA sequences}\footnote{An analogy for the removement, insertion and replacement of sequences can also be found in the biological DNA, where DNA sequences can be modified with the help of cutting enzymes or specific viruses.}. These operations allow the user to apply system updates at run-time.

\begin{figure}[ht!]
	\centering
		\includegraphics[width = .7\linewidth]{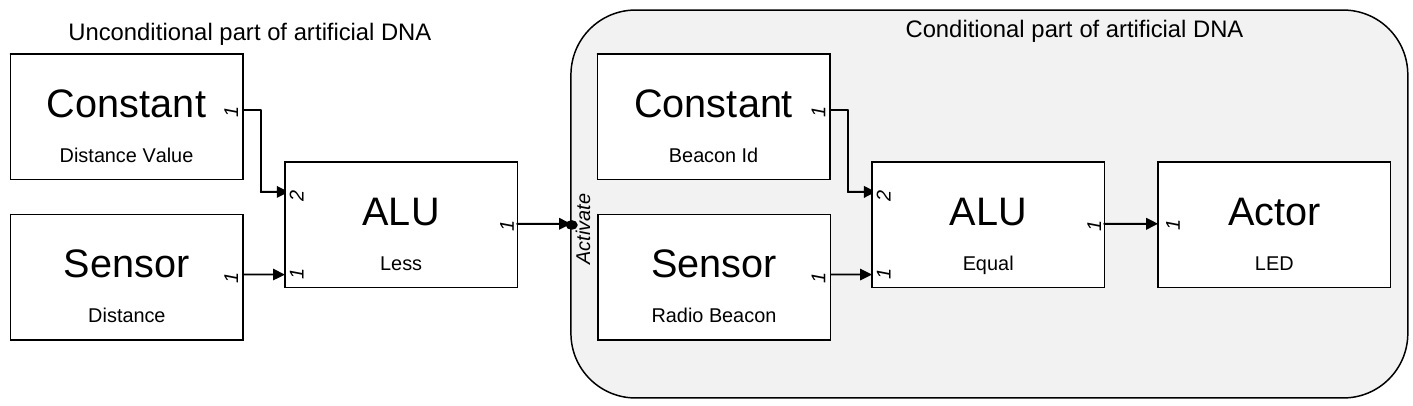}
	\caption{Example of a conditional artificial DNA}
	\label{fig:condDNA}
\end{figure}

If a reconfiguration of the artificial DNA is initiated by trigger conditions (e.g. given sensor values, points in time or calculated values), the application can react to changes in the environment. This directly leads to the concept of \textit{conditional artificial DNA}, where artificial DNA sequences are activated or deactivated by runtime events\footnote{This is more or less an analogy to gene regulation in the biological DNA, where gene expressions are activated or deactivated by inner or outer influences.}. Also parameter values of basic elements can be linked to trigger conditions. The trigger conditions will be calculated by the artificial DNA itself (e.g. a sensor value exceeds a threshold, a particular inner state is reached, a time span has elapsed, etc.). A simple example is shown in Figure \ref{fig:condDNA}. The left part of the artifical DNA is unconditional and therefore will be permanently built up. As soon as this part recognizes an object within a given distance, the conditional right part of the artificial DNA is activated and therefore will be built temporarily. This part checks if the object is a radio beacon with a predefined Id and activates a LED if this is true. As soon as the object leaves the given distance, the conditional part of the artificial DNA is deactivated and dismantled. According to the system architecture described in section \ref{sec:build}, building an artificial DNA sequence means all corresponding tasks (the basic elements) are allocated and interconnected. Dismanteling the sequence means all corresponding tasks and interconnections are completely deallocated and removed. So except from its rather small representation in the artificial DNA itself\footnote{Artificial DNAs are quite small ($< 1 kByte$), see e.g. Section \ref{sec:moreDNAs}. So the conditional part of an artificial DNA will usually consume only a few dozen bytes.} the conditional part does not consume any system resources (computation, memory\footnote{Since the basic elements in a conditional part are retrieved from the common generic basic element library like all other basic elements in the DNA, no additional code and data memory is necessary as long as the conditional part is not active.}, communication, energy) as long as it is not active. Therefore, the conditional artificial DNA enables a reconfiguration process similar to hardware reconfiguration, where several conditional parts can share the same system resources.

As next step, we will investigate and evaluate these reconfiguration mechanisms by including them into the prototypic implementation. An important research goal will be to determine the time complexity of the reconfiguration processes and the possibilities to keep real-time constraints during reconfiguration. A time complexity of $\mathcal{O}(n)$ with $n$ specifying the number of basic elements of the reconfigured artificial DNA sequence seems possible, since the building process of such a sequence can be performed in this time complexity (see Section \ref{sec:build}). Also, the potential of reconfiguring the artificial DNA in relation to its costs (implementation overhead, run-time overhead to trigger and execute a reconfiguration) will be examined.
Furthermore, reconfiguration is a pre-stage of evolution. All the operations necessary to reconfigure an artificial DNA by a user (insert, remove, replace sequences, modify parameters) are also basic operations for modifying the artficial DNA by an evolutionary process.  Additionally, conditional artificial DNA enables a specific and efficient variant of structural evolution: The user can predefine different alternative artificial DNA sequences for a particular part of the application. The system now can learn by an evolutionary process which alternative is the best in which situation and accordingly adapt the trigger condition. Furthermore, the alteratives themselves can be evolved. So situation dependend artficial DNA sequences can be generated. Since all alternatives share the same system resources (only the active one consumes resources, see above), the resulting overhead is minimized.

\subsection{Evolution}

While reconfiguration applies adaptions predefined by the user, evolution allows autonomous adaption by the system itself. The artificial DNA enables the use of evolutionary processes to modify, adapt and optimize the building plan of an embedded application at run-time. Generally, we can distinguish between \textit{parameter evolution} and \textit{structural evolution}. 

The parameter evolution leaves the structure of the building plan unchanged, but modifies parameters of the basic elements. Many basic elements have to be parameterized, e.g. a controller needs control parameters like P, I and D. The goal of parameter evolution is to optimize such parameters by using evolutionary principles.   
The parameters designated for optimization (usually a subset of the parameter set in the artificial DNA, e.g. the P, I and D parameters of balance control in the self-balancing vehicle) are initialized with starting values. Then, metrics and fitness functions are defined to judge the quality of application parts (e.g. the controller) or the entire application. Metrics for a controller can e.g. be transient response, settling time, accuracy, etc. As learning algorithms to optimize these parameters, evolutionary/genetic algorithms, but also other approaches like lerning classifier system can be used. The starting values are modified and optimized stepwise using the fitness functions.  
Interesting research questions are e.g. what kind of learning algorithms will work best for which type of parameters, how many steps will be necessary to improve the parameters and how different algorithms converge towards a better solution.

The structural evolution modifes the structure of the artificicial DNA itself. This means the used basic elements and their interconnections are changed by an evolutionary process. By defining rules for structural changes, the user can support, direct and guide this process. Such rules can e.g. be based on artifical DNA sequences and define which original sequences are allowed to be replaced by alternative sequences. By defining a pool of allowed alternative sequences, we can seemlessly move from reconfiguration to evolution. An example would be an orginal DNA sequence defining a PID controller with alternative DNA sequences for a fuzzy controller, an adaptive controller and a multistep controller. Using the above mentioned concept of conditional DNA, trigger conditions can be evolved to select one of the alternatives situation dependend.
Eventually, a library of artificial DNA sequences and possible alternatives can be created as a foundation of the structural evolution process. 
Other rules might allow random mutations for defined sequences or recombination of selected sequences. Combined with conditional DNA, also situtation dependent alternative sequences can be generated. To judge the quality of structural evolution, again the already mentioned metrics and fitness functions will be used. Interesting research questions are e.g. how the given optimization space affects time and outcome of the evolution process. Strict and narrow rules (only short sequences allowed for modification, only few alterantive sequences) constrain the process, while broad rules (long sequences allowed for modifications, many alternatives, mutation and recombination) allow also randomized modifications of the structure. The artificial DNA is an excellent enabler for structural evolution due to its linear characteristic, which maps easily to the coding methods of evolutionary algorithms or learning classifier systems. Also, the relationship between parameter evolution and structural evolution is an interesrting research question.

\begin{figure}[ht!]
	\centering
		\includegraphics[width = 0.8\linewidth]{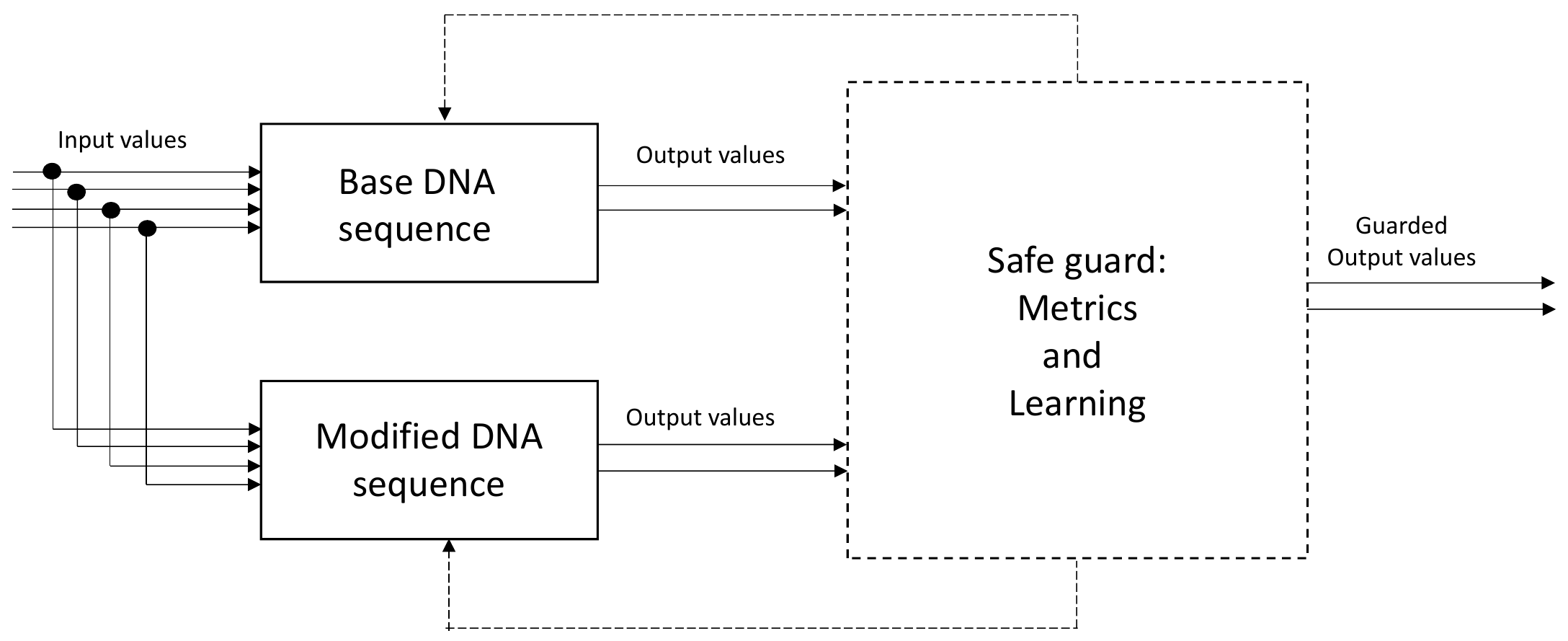}
	\caption{Evolution in real application environment}
	\label{fig:realevol}
\end{figure}

Finally, the process of evolution can be done in two different environments: the \textit{simulator} and the \textit{real application environment}. In the simulator, no special care has to be taken to prevent evolutionary steps from causing damage. It is a safe environment. The evolution can run in the simulated environment for as many steps as necessary and once the results are satisfying the resulting artificial DNA can be transfered to the real application. Running the evolution in the real application environment has the advantage of real interactions, environmental influences and timing conditions. However, evolutionary changes leading to a worse behavior could cause harm in the real environement. Figure \ref{fig:realevol} shows an approach to prevent this: an artificial DNA sequence to be optimized by evolutionary algorithms (regardless if parameter or structural evolution) is replicated. Only the copy is now target to the optimization process. The output values of the original sequence and the copy are surveilled by a safeguard unit. This unit uses the fitness functions and metrics necessary anyway to guide the evolution process. If the optimized sequence delivers better results than the original sequence, the output values of the optimized sequence are used. Otherwise, the output values of the orginal sequence are used and the evolution step is discared. Therefore, the original seqeuence is a fallback solution which prevents the system from damage. Furthermore, the original sequence can ensure to keep real-time constraints, if the optimizated sequence is not ready on time or delivers values too late. So the safeguard unit on the one hand represents a safety barrier and on the other hand is used for evolutionary progress. This unit can also be cascaded or structured hierachically.

Another issue of evolution in the real application environment is the safeguard unit has to run on the real hardware platform. Therefore, it should be kept as simple as possible to reduce the introduced overhead.  Evolution in the real application environment will use rather simple evolutionary methods (simple selections, mutations, crossovers), learning classifier systems and fitness functions (e.g. greater/less comparisons). The additional tasks necessary for evolution and fitness calculation can be distributed to the processors of the hardware platform by the AHS. In the simulator, any arbitrarily complex optimization algorithm can be used. As a compromise, also a hybrid solution is possible. Here, the real artificial DNA system on the real hardware platform interacts with a simulated environment. This implies the same overhead and simple evolutionary methods since the real hardware platform is used. However, the damage prevention mechanisms can be mostly ommitted since the environment is simulated. Only possible damage to the processors and communication system of the hardware platform has to be regarded. It is also often possible to keep the real-time behavior close to the real environment if the environment can be simulated timely precise. Just the real interactions are lost. Table \ref{table:evcomp} gives a comparison\footnote{For the robot vehicle platform FOXI (see Section \ref{sec:evaluation}) we already have developed a physically and timely precise simulation environment. So together with the artificial DNA simulator, the prototypic artificial DNA implementation and the real robot vehicle we already have a base to conduct experiments with these combinations.}. 

\begin{table}[ht!]
	\centering
  \begin{tabular}{c | c || c | c | c | c | c }
Hardware & Application & Overhead in & Evolutionary & Damage     & Real-time & Real\\
Platform & Environment & Hardware Platform & Methods      & Prevention &  Behavior & Interactions\\
\hline       
Simulator & Simulator & No  & Complex & No & No & No\\
Real      & Simulator & Yes & Simple & Mostly No & Mostly Yes & No\\
Real      & Real      & Yes & Simple & Yes & Yes & Yes\\
  \end{tabular}
	\caption{Posibilities and properties of evolution}
	\label{table:evcomp}
\end{table}

Interesting research questions are how big the overhead introduced by evolution in the real hardware platform will be and if this pays compared to the pure simulator solution? How precise the environment has to be simulated for the hybrid solution? To what extend evolution can be done with damage prevention in the real application environment? To what extend real-time constraints can be kept there?

Overall, the artificial DNA represents a key enabler to introduce and evaluate well known evolutionary methods to the field of embedded real-time systems for evolution at run-time. In traditional approaches, evolutionary or genetic methods are mainly used to solve complex non-realtime optimization and matching problems (like e.g. the travelling salesman problem, the knapsack problem, find functions matching sequences of values, and many more.). Here, evolution is used to build and optimize embedded systems at run-time.

\subsection{Run-time Synthesis of Fault Tolerance Mechanisms}

The artificial DNA has the inherent capability of self-healing core-failures\footnote{The complete breakdown of a core.} as described in Section \ref{sec:basicidea} and evaluated in section \ref{sec:evaluation}. Additionally, it enables the adaptive synthesis of further fault tolerance mechanisms at run-time. This will be investigated by future work. First, the user can define importance values for particular actors (or other basic elements) in an artificial DNA. This allows to distinguish more important parts from less important parts (e.g. the brake actors in a car are more important than the brake light actors or the window lift actors. In the automotive area, this is reflected by ASIL (Automotive Safety Integrety Levels)). Starting with the given importance values, the system can now backtrack the artificial DNA back to the inputs and assign each basic element a resulting importance value. This has two major advantages: First, the system can now automatically discard the less important basic elements and thereby keep the more important parts of the artificial DNA active and running if the core redundancy is depleted by too many core failures. The available hardware ressources are exploited the best possible way. Second, depending on the importance values and available resources (number and state of the available processor cores) the system can automatically and dynamically replicate important parts of the artificial DNA at run-time (for double or triple modular redundancy) to identify and remove errors and error sources (e.g. unreliable processor cores). Highly important artificial DNA sequences (e.g. ABS or steer by wire in a car) can be replicated at run-time to detect errors. In case of errors and enough cores available, a third instance can be dynamically created to correct errors and possibly spot an unreliable core. Since the artificial DNA is present on all processor cores, replicas can be flexibly built and relocated at need and available resources. 

\begin{figure}[ht!]
	\centering
		\includegraphics[width = 0.8\linewidth]{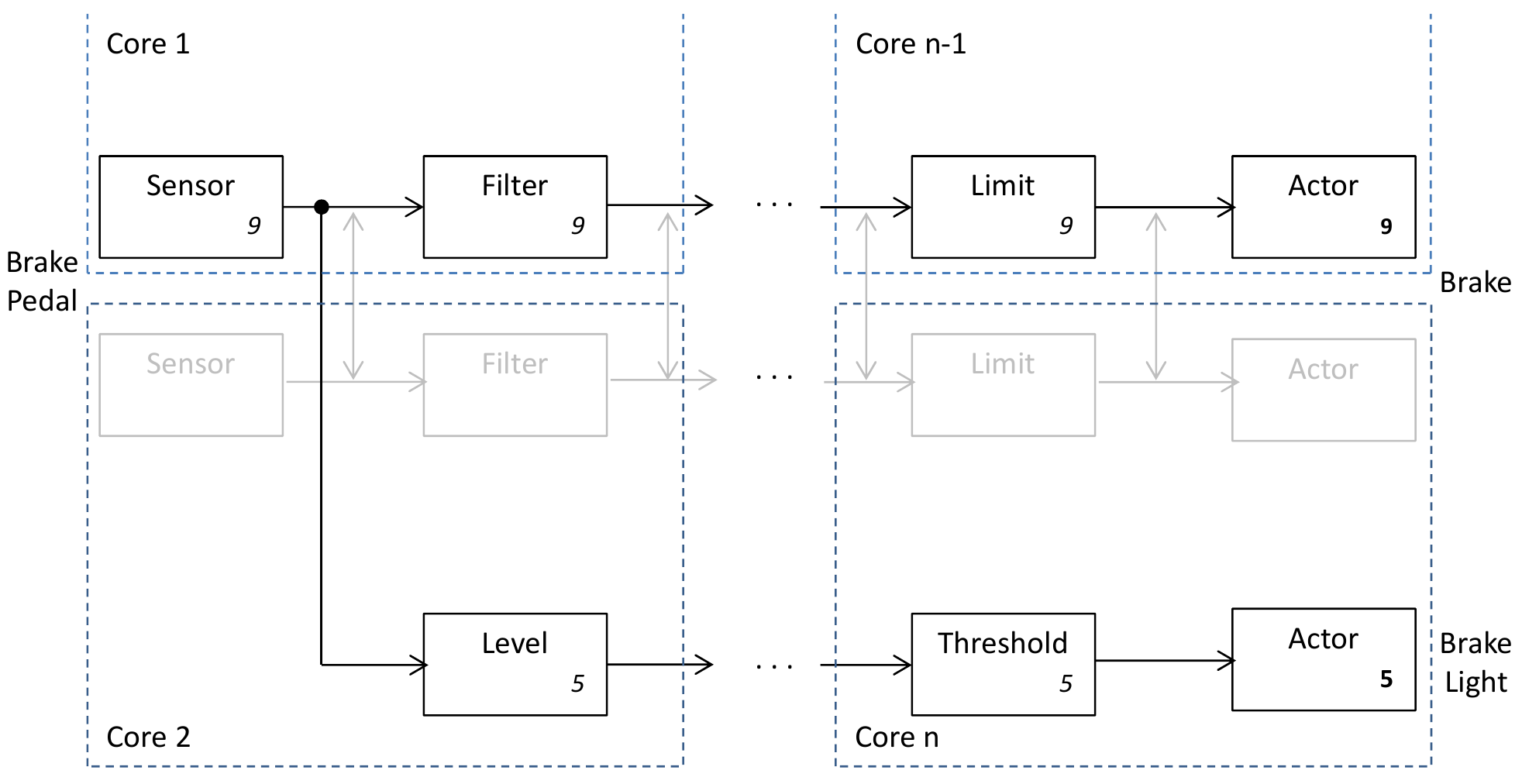}
	\caption{Dynamic synthesis of modular redundancy for basic elements based on importance values}
	\label{fig:synthred}
\end{figure}

Figure \ref{fig:synthred} gives a simplified example for an artificial DNA controlling the brakes and the brake lights of a car. Here, the brake actor has an importance value of 9 while the brake light actor is less important and therefore has an importance value of only 5. Starting with the actors, the importance values are backtracked to the sensors (the brake pedal sensor in this example\footnote{Other sensors important for the braking system like e.g. wheel sensors are not shown in this simplified example to not overload the picture, however they are handled exactly the same way.}). Based on the artficial DNA, the system has now automatically replicated all basic elements with an importance value of 9 and allocated them to cores. To be effective in detecting temporary or permanent core errors, replicas have to be allocated to different cores than the original basic elements. This can be easily achieved by the AHS. Also diversity can be considered by trying to put replicas on heterogeneous cores. Another possibility is to provide diverse implementations for basic elements in the basic element library. If available, a different implementation could then be selected as replica to spot implementation errors. Special care has to be taken when replicating sensor and actor basic elements. These elements provide the access to the physical sensors and actors. Just replicating them with access to the same physical sensor or actor would only allow to detect and handle errors in sensor and actor to core communication. If replicas of the physical sensors and actors are available, a replicated basic element can automatically connect itself to a replicated sensor or actor. This allows to also handle errors of sensors and actors themselves. 

The artifical DNA enables a new flexible and dynamic use of the well-known concept of modular redundancy at run-time. Replicas can easily share cores with basic elements from other parts of the artficial DNA, as shown in Figure \ref{fig:synthred}. Replicas can relocate themselves at run-time, remove themselves if resources are getting low or add themselves if more resources become available. This is different from traditional static modular redundancy. If the artificial DNA is changing by reconfiguration or evolution (see previous Sections) the replicas adapt automatically. Since each processor core knows the current building plan of the entire system, modular redundancy can construct or reform itself autonomously according to the current situation. Location, amount, selection and structure of redundant parts are self-adaptive. Therefore, it is also different from dynamic redundancy approaches like e.g. proposed in projects like \cite{DFG-SPP-89}.
  
Since replicas are also used for evolution in real environment (see above), synergies can be exploited here. Main research questions are how to define the relationship between importance values, available resources and replicas generated? How efficient and fast the replicas can be created and removed, how results are efficiently compared and if modular redundancy should be better applied per basic element or per sequence?

Instead of replicating entire artificial DNA sequences, just a single or even a partial sequence could be selected for replication at a given time slot. This selection changes over time, so that always a small but varying part of the artificial DNA is replicated. A round robin scheme weighted by the importance values could e.g. be used to change the selection. Thereby more important parts are replicated more frequently than less important parts. This technique does not permanently spot errors in critical data paths like the approach above. However, possible unreliable or failing locations (cores) for basic elements can be located. The detection probability increases with the importance value.

Another possibility of extended fault tolerance is using the implicit redundancy given by an artificial DNA. Since an artificial DNA consists of rather simple basic elements, most of these elements will occur multiple times in an artificial DNA (e.g. an arithmetic/logic unit, a PID controller, ...). These multiple occuring elements could check themselves mutually, e.g. by exchanging challences when they are not busy. This technique also allows to detect unreliable or failing locations for basic elements with minimum overhead.

Extending self-healing mechanisms is another interesting and challenging research field. As could be seen the artificial DNA can be exploited in many ways to increase fault-tolerance. We consider it being worth for further investigation and examination in our future work.

\section{Conclusions}
\label{sec:conclusions}
In this technical report we presented in detail an approach to use a digital artificial DNA for self-describing and self-building systems. This DNA is deposited in each computation node as a blueprint to build, adapt and repair the system autonomously at runtime. Mimicking biology this way provides robustness and dependability. The prototypic implementation of the DNA approach enabled an evaluation on a real-world scenario, a robot vehicle. The results showed that the approach is feasible and promising. The memory and communication overhead of the implementation are rather small, application DNAs are compact and can be built with a limited number of basic elements. Self-building, -adapting and -repairing of the application at runtime can meet real-time requirements. 

The future work we described will focus on dynamic DNA modification by reconfiguration and evolution as well as investigating the possiblities to synthesize additional fault-tolerance mechanisms dynamically and automatically at run-time. The artificial DNA enables the introduction of these concepts to embedded real-time systems.

\bibliographystyle{plainnat} 
\bibliography{LitDB_ES} 

\end{document}